\documentclass[preprint,amsmath,amssymb,nofootinbib,floatfix]{revtex4}
\usepackage{graphicx}
\renewcommand{\section}{\arabic{section}}
\begin{document}



\title{Synchronicity From Synchronized Chaos}

\author{Gregory S. Duane\protect\footnote{corresponding author: 
email address gregory.duane@colorado.edu}}

\address{Dept. of Atmospheric and Oceanic Sciences\\
UCB 311\\University of Colorado\\ Boulder, C0 80309-0311\\ 
and\\Macedonian Academy of Sciences and Arts\\
Bul. Krste Misirkov 2, P.O. Box 428\\ 1000 Skopje, Republic of Macedonia}

\date{\today}


\newpage

\begin{abstract}
The synchronization of loosely coupled chaotic oscillators, a phenomenon 
investigated intensively
for the last two decades, may realize the philosophical notion of  
``synchronicity". Effectively unpredictable
chaotic systems, coupled through only a few variables, commonly exhibit a 
predictable relationship
that can be highly intermittent.  We argue that the phenomenon closely resembles 
the notion of meaningful  synchronicity put forward by Jung and Pauli if one 
identifies ``meaningfulness" with
internal synchronization, since the latter seems necessary for 
synchronizability with an external system.  Jungian synchronization of mind and 
matter is realized if mind is analogized to a computer model, synchronizing with 
a sporadically observed system as in meteorological data assimilation.
Internal synchronization provides a recipe for combining different models of the 
same objective process, a configuration that may also describe the functioning 
of conscious brains.  In contrast to Pauli's view, recent developments suggest a 
{\it materialist} picture of semi-autonomous mind, existing alongside the
observed world, with both exhibiting a synchronistic order.  Basic physical 
synchronicity is manifest in the
non-local quantum connections implied by Bell's theorem. The quantum world 
resides on a
generalized synchronization ``manifold", a view that provides a bridge between 
 nonlocal  realist interpretations and local realist interpretations that 
 constrain observer choice .
 \end{abstract}



\maketitle


\section{Introduction}

\label{sec:intro}

Synchronization within networks of oscillators is widespread in nature, even
where the mechanisms connecting the oscillators are not immediately apparent. One recalls the example
 of the synchronization of clocks                            
suspended on a common rigid wall, a paradigm commonly attributed to Huygens. As with similar phenomena of fireflies blinking in
unison, or female roommates synchronizing hormonal cycles, the pattern suggests a univerally valid organizational
principle that transcends any detailed causal explanation. Further from everyday experience, but perhaps
related \cite{Dua05}, are the quantum mechanical harmonies between distant parts of a system that are not causally 
connected,  involving also the observer's choice of measurements as implied by
Bell's Theorem \cite{Bell}.

The study of coupled networks of oscillators in classical physics has focussed 
on regular oscillators with periodic
limit-cycle attractors. Such models afford explanations for such surprising 
relationships as the one observed
by Huygens, but other synchronous relationships that are sometimes said to exist
in nature are less easily explained.  While the 
synchronization of chaotic oscillators with strange attractors has become familiar
in the last two decades, most work on such systems has examined engineered systems, primarily for application
to secure communications, using the low-dimensional signal connecting the oscillators as a carrier that is
difficult to distinguish from noise.  However, a examples of synchronized chaos in pairs of systems of
partial differential equations that describe physical systems, coupled loosely, 
have also been given \cite{Koc,Dua97,Dua99,Dua01,Dua04}.

In the philosophical realm, synchronous relationships 
that are difficult to explain causally have figured  prominently in primitive 
cultures and in traditions commonly associated with Jung\footnote{No reference is made in this paper to the use of archetypes
in physical theory or other aspects of Jung's philosophy.}.\cite{Peat87}  The 
notion of ``synchronicity" commonly associated
with Jung has two essential characteristics beyond the simple 
simultaneous occurrence of corresponding events:
First, the simultaneous occurrences or ``synchronicities" must be isolated 
occurrences. Second, the synchronicities
must be ``meaningful". The idea of synchronicity thus goes beyond the synchronization of oscillators in positing a
new kind of order in the natural world, schematized by Jung and Pauli (Fig.
 \ref{figJung}) in
their book {\it The Interpretation of Nature and of the Psyche} \cite{Jung}.  
Regular oscillator models fall far short of explaining  synchronicities of this type,
as Strogatz observed in his popular exposition \cite{Str}. A
particularly important instance is the synchronization of matter and
mind. In this view, mind is not slaved to the objective world, but tends
to synchronize with it, based on limited exchange of information.  Jung's
examples of synchronicity, and subjective perceptions of synchronicity generally,
are often dismissed as the result of chance, but a minority opinion
follows Pauli in asserting that a synchronistic order
exists in the world alongside the causal one.

While Pauli kept such speculations largely separate 
from his scientific work, the point of this paper, reached through a review of the author's and others'
more recent work, is to show that the nonlinear dynamics paradigm of synchronization in networks of loosely
coupled chaotic systems can realize the philosophical notion of synchronicity,
or at least approach it much more closely than is possible with regular
oscillators. The proposed realization is concrete in nature, without any need
for dualism between the mental and material worlds. It is also different from
the ``dual-aspect monism" ascribed to Jung and Pauli themselves \cite{HA},
in that material synchronization is put forward as an explanation of synchronous
relationships in the mental realm, and between mind and matter.  

We begin by showing, in 
the next section, that the simple
introduction of a time delay in the coupling between the systems can transform 
a situation of complete synchronization
to one of isolated ``synchronicities".  In Section \ref{sec:percept}, we review previous work on an application of synchronized
chaos to ``data assimilation" of observations of a ``real" system into a computational model that is intended to
synchronize with truth, analogously to the synchronization of matter and mind.  ``Meaningfulness" is naturally interpeted
as internal coherence.  A three-way relationship between two parts of a real system
and a third system conceived as an observer is shown to satisfy the requirement
for meaningfulness in synchronicities in Section \ref{sec:meaning}.

The objective rational basis for synchronicity that is put forward in this 
paper suggests applications of the new
organizational principle to 
processes in the brain and in the physical world. In Section \ref{sec:comp}, we discuss 
implications of synchronized chaos for neural systems,
in view of contemporary ideas about synchronization as a binding mechanism in 
perception and consciousness. In
Section \ref{sec:quantum}, we argue that synchronized chaos can support quantum nonlocality and the Bell correlations in a
realist interpretation, where determinism is retained.  The concluding section 
speculates on remaining gaps
between our objective realization of the synchronicity principle and its 
original philosophical motivation.
  
\section{Highly Intermittent Synchronization in Loosely Coupled Chaotic Systems}

\label{sec:back}
Extensive interest in synchronized chaotic systems was spurred by the work
of Pecora and Carroll \cite{Pec} (cf. \cite{Fuj,Afr}), who initially studied the
phenomenon in 3-variable Lorenz systems \cite{Lor63}, a prototypical example of
chaos.  Slaving the $X$ variable of one Lorenz system to the corresponding
variable of a second system, one has:

\begin{minipage}[t]{3in}
\begin{eqnarray*}
\hspace{.4in}\dot{X}&=&\sigma(Y-X)  \nonumber \\
\dot{Y}&=&\rho X - Y - X Z   \nonumber \\
\dot{Z}&=&-\beta Z + X Y   \nonumber\\
\end{eqnarray*}
\end{minipage}
\begin{minipage}[t]{3.5in}
\begin{eqnarray}
\label{eqPC}
&& \nonumber \\
\dot{Y}_1&=&\rho X - Y_1 - X Z_1    \\
\dot{Z}_1&=&-\beta Z_1 + X Y_1 \nonumber
\end{eqnarray}
\end{minipage}\\
The systems 
defined by $(X,Y,Z)$ and $(X_1=X,Y_1,Z_1)$, respectively,  synchronize rapidly: As $t\rightarrow \infty$, $Y_1(t)-Y(t)\rightarrow
0$, $Z_1(t)-Z(t)\rightarrow 0$, as shown in Fig. \ref{figPC}. (Synchronization also occurs if the slave system is
driven by the master $Y$ variable instead of the $X$ variable, but not if
driven by the $Z$ variable.) Various schemes to use chaos synchronization for
cryptography were motivated by the thought that variables analogous to $X$ in
(\ref{eqPC}) could be used as carrier signals that would be difficult to 
distinguish from noise \cite{PecReview} - the signal between the two systems defined by X 
is broadband and has no characteristic frequency.   Takens Theorem \cite{Takens} 
can still be used to infer information about the encoding from a segment of 
signal that is sufficiently long,  but as one considers higher-dimensional 
analogues of (1),  it becomes increasingly difficult to do so.

The two connected chaotic systems are conceived as effectively 
unpredictable, but exhibit significant correlations if connected by a 
signal that is conceived as effectively inscrutable.  Taking these views 
of the systems and their connection as valid, we have an instance of an 
acausal synchronous relationship.  To 
the extent that relationships between physical systems analogous to (\ref{eqPC}) 
occur in nature, synchronism becomes a valid physical principle.

It is important that the ``acausal"  correlations arise in the context of
a perfectly causal, deterministic system.  But the same could be said of the
examples of synchronicity given by Jung. Those surprising coincidences occurred
in systems that one would imagine to be governed by ordinary deterministic
physics, with a history of connection between the two systems, but that
connection would not have been readily interpreted as causal.  The synchronistic
behavior emerges as a higher-order relationship in causal systems that could
not have been predicted from existing causal theory.  Synchronicity thus does not logically
contradict causality, but transcends it. If physical systems manifest such
relationships as in the pair of coupled Lorenz systems, as will be illustrated
below, then synchronicity might be accounted for rationally.

Synchronization can indeed occur with weaker forms of coupling than the complete
replacement of one variable by its corresponding variable as in (\ref{eqPC}),
but degrades below a threshold coupling strength. Typically, synchronization
degrades via on-off intermittency \cite{Ott}, where bursts of desynchronization
occur at irregular intervals, or as ``generalized" synchronization
\cite{Rul},
where a strict correspondence remains between the two systems, but that
correspondence is given by a less tractable function than the identity.
As shown schematically in Fig. \ref{figgensync}, as differences between
the two systems increases, the correspondence changes from the identity to a smooth function
that approximates the identity, to one given by a function that is nowhere
differentiable. The last case is in fact common \cite{So}.

While initial research on synchronized chaos was
motivated by potential applications to secure communication schemes using 
electronic circuits, the phenomenon has also been demonstrated in lasers
\cite{Roy} 
and ferromagnetic materials \cite{Peterman}, as well as in the fluid dynamical systems 
discussed below. In applications to
physical
systems it is natural to consider forms of coupling that embody a time
delay.
If one extends chaos synchronization to the realm of naturally occurring 
systems, the delay in transmission ought to be described in terms of the same
physics that governs the evolution of the systems separately. To a
first approximation let us assume that the time scale of the delay is
the same as some intrinsic dynamical time scale of each system. Consider
the following configuration of two Lorenz systems, coupled through an
auxiliary variable  $S$  that introduces a delay:
  
\begin{minipage}[t]{3in}
\begin{eqnarray*}
\hspace{.2in}\dot{X}&=&\sigma(Y-X) \nonumber\\
\dot{Y}&=&\rho (X-S) - Y - (X-S) Z   \nonumber \\
\dot{Z}&=&-\beta Z + (X-S) Y   \nonumber\\
\dot{S}&=&-\Gamma S + \Gamma (X-X_1)\nonumber \\
\end{eqnarray*}
\end{minipage}
\begin{minipage}[t]{3in}
\begin{eqnarray}
\label{PCdelay}
\dot{X}_1&=&\sigma(Y_1 - X_1) \nonumber\\
\dot{Y}_1&=&\rho (X_1 + S) - Y_1 - (X_1 + S) Z_1\nonumber \\
\dot{Z}_1&=&-\beta Z_1 + (X_1 + S) Y_1 
\end{eqnarray}
\end{minipage}\\
The system (\ref{PCdelay}) is a generalization of the Pecora-Carroll coupling
scheme (\ref{eqPC}) to a case with bidirectional coupling and
where each subsystem is partially driven and partially autonomous.

As $\Gamma\rightarrow \infty$ in (\ref{PCdelay}), with $\dot{S}$ finite,
$S \rightarrow X-X_1$. In this limit, the system reduces to 
a bidirectionally coupled version of (\ref{eqPC}),
which indeed synchronizes. In the general case of the coupled system
(\ref{PCdelay})  with finite
$\Gamma$, the subsystems exchange information more slowly: if $X$
and $X_1$ are slowly varying, then S asymptotes to $X-X_1$ 
over a time scale $1/\Gamma$. Thus $\Gamma$ is an inverse time lag in the 
coupling dynamics.

Synchronization along
trajectories of the system
(\ref{PCdelay}) is represented in Fig. \ref{figsyncdelay} as the 
difference $Z-Z_1$ vs. time, for  decreasing values of $\Gamma$. For large
$\Gamma$, the case represented in Fig. \ref{figsyncdelay}a, the subsystems 
synchronize.
As $\Gamma$ is decreased in Figs. \ref{figsyncdelay}b-d,
 corresponding to
increased time lag, increasingly frequent bursts of desynchronization
are observed, until in Fig. \ref{figsyncdelay}d (uncoupled systems) no portion of the
trajectory is synchronized. The bursting behavior can be understood
as an instance of on-off intermittency \cite{Pla},\cite{Ott}, the
phenomenon that can occur when an invariant manifold
containing an attractor loses stability, so that the attractor is
no longer an attractor for the entire phase space, but is still
effective in portions of the phase space. Trajectories then
spend finite periods very close to the invariant manifold, interspersed
with bursts away from it.

The case of a coupling time lag that is of the same order as the prescribed physical 
time scale in
the simple Lorenz system corresponds to $\Gamma=1$, with behavior as in
Fig. \ref{figsyncdelay}c. Although there is little trace of synchronization,
the average instantaneous distance between the
subsystems is less than it is in the uncoupled case. More interestingly, there
is a very short period of nearly complete synchronization. In a very long
integration, such
"synchronicities", of moderately short duration, occur more commonly than they
would by chance in unrelated systems, as seen in the histograms in Fig.
\ref{figsyncdelay}e, showing total time in synchronicities of given
duration for the two cases. 

The system
(\ref{PCdelay}) is indeed analogous to one derived from a pair of geophysical
fluid models coupled by standing waves in narrow ducts
\cite{Dua97}. Auxiliary variables analogous to $S$ in
(\ref{PCdelay}) arise by first decomposing the field into a piece that satisfies the 
full nonlinear equations with homogeneous boundary conditions and
a second piece that satisfies a linear system with matching boundary conditions 
in the region of the narrow ducts. The linear equations are solved using boundary
Green's functions that effect a time delay. The auxiliary variables are integrals of 
products of the boundary Green's functions and differences
of corresponding field variables from the two sides of the ducts.
Intermittent synchronization of the two ODE systems implies correlations between 
large-scale
weather patterns in the midlatitude regions of the Northern and Southern 
Hemispheres, since they are connected by ``duct" regions in the 
tropics where prevailing winds are westerly and Rossby waves can thus 
propagate \cite{Dua97}.

The co-occurence of the weather patterns in the two hemispheres could not have been
predicted from the mere existence of the narrow connecting ducts. One needs to study
the real system, or sufficiently complex numerical model thereof, to observe the
synchronous behaviour.  The small statistical correlations induced by wave
transmission through the ducts were indeed surprising 
(but
not particularly useful), although the ``blocking" behavior, to be described in 
Section \ref{sec:meaning}  had been familiar for several decades. 
The tendency of blocks to co-occur, which is reminiscent of Huygens' clocks or
synchronized hormonal cycles in female roommates, but constructed from systems
that are individually unpredictable, supports the existence of synchronicity as
an organizational principle that transcends causality.

The impact of  chaos synchronization is enhanced 
greatly  by  the phenomenon of small-world  networks.    One can consider a 
large array of chaotic oscillators with local and/or long-range connections 
among them.   In a small-world network,  a few long-range connections are 
sufficient to sharply decrease the average degree of separation between any 
two oscillators, as occurs more generally in a scale-free network in 
which the number of highly connected nodes decreases
with the number of their connections according to a power law \cite{Str}.  
Randomly connected networks of this type can be expected to synchronize more readily
than regular networks that are connected only in local neighborhoods: 
 the introduction of a few long-range connections
can lead to a phase transition to long-range synchronization.
\cite{Lag},\cite{Bar},\cite{Hua},\cite{Zha}.

\section{Machine Perception as Chaos Synchronization}

\label{sec:percept}

The connection between synchronized chaos and mind-matter synchronicity is best
illustrated by another application to meteorology, a field that inspired the 
modern notion of chaos \cite{Lor63}.
Computational models that predict weather include a feature not found in numerical
solutions of simpler initial-value problems: As new data is provided by observational
instruments, the models are continually re-initialized. This {\it data
assimilation} procedure combines observations with the model's prior prediction
of the current state - since neither observations nor model forecasts are 
completely reliable - so as to form an optimal estimate of reality at each instant
in time.  While similar problems exist in other fields, ranging from financial
modelling to factory automation to the real-time modelling of biological or
ecological systems, data assimilation methods are more developed in 
meteorology than in other fields.  

Since the problem of data assimilation arises in any situation requiring
a computational model of a parallel physical process to track that process
as accurately as possible based on limited input, it has been asserted that
the broadest view of data assimilation is that of machine perception by an
artificially intelligent system \cite{Dua06}.  Like a data
assimilation system, the human mind forms a model of reality that functions
well, despite limited sensory input, and one would like to impart such an
ability to the computational model. 

The usual approach to data assimilation is to regard it as a tracking problem
that can be solved using Kalman filtering or generalizations thereof. But
clearly the goal of any data assimilation is to synchronize model with reality,
i.e. to arrange for the former to converge to the latter over time.  Thus
the synchronously coupled systems of the previous section are re-interpreted
as a ``real" system and its model. In the system (\ref{eqPC}), for instance,
we imagine that the world is a Lorenz system, that only the variable $X$ is
observed, and that the observed values are passed to a perfect model.  

The above philosophical considerations have motivated an attempt to recast data 
assimilation as a synchronization problem and thus to improve assimilation 
algorithms \cite{Dua06,Abarb,Yang}. It may seem a large stretch from 3-variable
systems to the human mind, but the use of low-order systems to study problems
that arise in numerical weather models is common practice in meteorology, 
popularized especially by E. Lorenz.  These numerical models, intended to
represent turbulent fluid behaviour over a vast range of scales, are among the 
most complex computational models known.  It is remarkable that such models are
capable of tracking reality using only a sparse, temporally intermittent set of
noisy observations. To demonstrate the relevance of
synchronization in low-order systems, it is first necessary to show that the 
phenomenon 
persists as 
the dynamical dimension of the model is increased to realistic values.
Chaos synchronization in the sort of
models given by systems of partial differential equations that are of interest
in meteorology and other complex modelling situations has indeed been established.
Pairs of 1D PDE systems of various types, coupled
diffusively at discrete points in space and time, were shown to synchronize
by Kocarev {\it et al.} \cite{Koc}. 

Synchronization in geofluid models that are relevant to weather prediction was
demonstrated by Duane and Tribbia \cite{Dua01},\cite{Dua04}.  The models \cite{Vau} are given in terms
of the streamfunction $\psi(x,y,i,t)$ in a 2-layer ($i=1,2$) channel, contours of which are streamlines of 
atmospheric flow, as shown in Fig. \ref{figQGsync}, and a derived field
variable, the potential vorticity $q_i\equiv f_0 + \beta y + \nabla^2 \psi_i + 
R_i^{-2} (\psi_1-\psi_2) (-1)^i$, with constants as defined in the reference.  
The dynamical equation for each model is  
\begin{equation}
\label{QGmodel}
Dq_i/Dt\equiv
\partial q_i/\partial t+J(\psi_i,q_i)=F_i+D_i
\end{equation}
where the Jacobian
$J(\psi,\cdot)\equiv \frac {\partial \psi}{\partial x}\frac{\partial \cdot}
{\partial y}
-\frac {\partial \psi}{\partial y}\frac{\partial \cdot}{\partial x}=
v_y\frac{\partial \cdot}{\partial y}
+v_x\frac{\partial \cdot}{\partial x}$ gives 
the
advective contribution to the Lagrangian derivative $D/Dt$.
The equation (\ref{QGmodel}) states that potential vorticity is conserved on a moving
parcel,
except for forcing $F_i \equiv \mu (q^*_i-q_i)$ and dissipation $D_i$ as defined
by
Duane and Tribbia \cite{Dua04}. The forcing induces a relaxation to a jet-like
background flow $\psi^*$ (Fig. \ref{figQGsync}a,b) with $q^*\equiv q(\psi^*)$,
injecting energy into the system.

Two models of the form (\ref{QGmodel}), ${D q^A}/{D t}=F^A+D^A$ and
${D q^B}/{D t}=F^B+D^B$
were coupled diffusively through a modified forcing term
\newcommand{\bk}{{\bf k}}
$F^B_\bk=\mu^c_\bk[q^A_\bk-q^B_\bk] + \mu^{ext}_\bk  [q^*_\bk - q^B_\bk]$,
where the flow has been decomposed spectrally and the subscript $\bk$ on each
quantity indicates the wave number $\bk$ spectral component.   The two sets of coefficients $\mu^c_\bk$ and 
$\mu^{ext}_\bk$ were chosen to couple the two channels only in some medium range
of wavenumbers.

It was found that the two channels rapidly synchronize if only the medium 
scale
modes are coupled (Fig. \ref{figQGsync}), starting from different initial flow
patterns. For unidirectional coupling,  the synchronization would
effect assimilation of Fourier-space data from the $A$ channel into the $B$ 
channel. The restriction to coupling of medium-scale modes is a Fourier-space
counterpart to assimilating data from discrete, well-separated observation
points.
It has been shown analytically that optimal
synchronization is
equivalent to Kalman filtering when the dynamics change slowly in phase
space, so that the same linear approximation is valid at each point in time
for the real dynamical system and its model. When the dynamics change rapidly,
as in the vicinity of a regime transition, one must consider the full
nonlinear
equations and there are better synchronization strategies than  Kalman 
filtering or the popular method of ensemble Kalman filtering \cite{EnKal}. The deficiencies of the
standard methods, which
are
well known in such situations, are usually remedied by ad hoc corrections,
such as ``covariance inflation" \cite{And}.
In the synchronization view, such  corrections can be derived from first
principles \cite{Dua06, DuaSpringer}, demonstrating the 
analytical power of the perception-as-synchronization-of-truth-and-model 
viewpoint.  

\section{Internal Sync vs. Mind-Matter Sync and the Role of Meaning}
\label{sec:meaning}

In the context of data-asssimilation/machine-perception, the role of synchronism is indeed
reminiscent of Jung's notion of synchronicity in the relationship between mind 
and the material world. But in the latter view, and in the popular culture
surrounding the notion, the alleged relationships between events, mental or physical, 
are detected without close inspection and are ``meaningful". In this section, 
it is argued
that meaningfulness is realized naturally in terms of the internal coherence
that is typically present in any system that synchronizes with an external
system, and thus that the scientific view of synchronization should 
satisfy philosophers in this regard as well.

Prior use of the idealized
geophysical model considered illustrates how ``meaning" would enter. The
quasigeostrophic channel model was originally developed to represent the
geophysical ``index cycle", in which the large-scale mid-latitude 
atmospheric circulation vacillates, at apparently random intervals, between two types
of flow \cite{Vau}.  In the ``blocked flow" regime, e.g. Fig. \ref{figblockzon}a,
a large high-pressure center, typically over the Pacific or Atlantic, interrupts
the normal flow of weather from west to east and causes a build-up of extreme
conditions (droughts, floods, extreme temperatures) downstream. In the ``zonal flow"
regime, e.g. Fig. \ref{figblockzon}b, weather patterns progress normally.  Synchronization of flow states,
complete or partial, implies correlations between the regimes occupied by two
coupled channel models at any given time. Such correlations, in the context of synchronization between reality and model, are indeed
meaningful to meteorologists and to the residents of the regions downstream of
any blocks. Synchronization of two highly simplified versions of the channel
model has been used to predict correlations between blocking events in the
Northern and Southern hemispheres \cite{Dua97,Dua99}, and
synchronization  of two channel models has been used to infer conditions under
which Atlantic and Pacific blocking events can be expected to anticorrelate
\cite{Dua01,Dua04}.

To generalize from the
geophysical models, we note that blocks are ``coherent structures", as commonly
arise in a variety of nonlinear field theories. Such structures, of which
solitons are perhaps the best known example, persist over a period of time because
of a balance between nonlinear and dispersive effects. 
While no generally accepted definition of ``coherent structure" has been
articulated, one view of their fundamental nature can support the proposed
general connection with meaningfulness. For a structure to persist, the
different degrees of freedom of the underlying field theory must continue
to satisfy a fixed relationship as they evolve separately. That behavior
defines generalized synchronization, the phenomenon
in which two dynamical systems synchronize, but with a correspondence
between states given by a relationship other than the identity \cite{Rul}.
Coherent structures are then characterized by {\it internal} generalized 
synchronization within a system.  As state variables that are generally
synchronized with other state variables reveal additional information, it
is proposed that such relationships capture ``meaningfulness" in the usual 
sense of that term.

Meaningfulness is even more naturally defined as internal synchronization
within mind. A response to a given external stimulus by any ``element" of mind
is likely to be deemed meaningful if there are synchronized, parallel responses 
of other mental elements.

No reference has been made to the semantic {\it meaning} of the meaningful
structures. Rather, the function of internal synchronization is that of parsing, or
perceptual grouping, which must precede interpretaion. Synchronization is indeed
known
to play a role in perceptual grouping in real neural systems, as discussed below in Section
\ref{sec:comp}.  This leaves the question of whether the correspondence between
two co-occurring structures is itself meaningful, i.e. whether the structures
are properly paired, but in the typical case (especially in the case of a real
system and its model), there is enough similarity between the synchronizing
systems that a meaningful pairing is likely.
   
It remains to show that internal synchronization is
likely in each of a pair of dynamical systems that exhibit synchronized chaos.
It has indeed been hypothesized that internal synchronization is {\it required} for 
synchronizibility with an external system \cite{Dua09}.
The essential role of coherence in synchronizing systems was highlighted
by considering a pair of Hamiltonian systems, for which complete synchronization
is precluded because phase-space volumes of ensembles of trajectories are
preserved, by Liouville's Theorem.   We consider a nonlinear scalar field theory that 
gives rise to ``oscillons" - coherent structures in the
field that
oscillate in fixed, randomly placed locations - as do similar structures that
were first noted
in vibrating piles of sand \cite{Umb}.   The expansion of the universe
plays a role in the cosmological case that is analogous to the role of
frictional dissipation in the sandpiles, but the system is governed by
a time-dependent Hamiltonian, and Liouville's Theorem still applies.  A
one-dimensional model
is given by the relativistic scalar field equation, with a nonlinear potential
term, in an expanding background geometry described by a Robertson-Walker 
metric with Hubble constant $H$.  Using covariant derivatives for that
metric in place of ordinary derivatives, one obtains the field equation
\begin{equation}
\label{osc}
\partial^2 \phi/\partial t^2+H\partial \phi/\partial
t-e^{-2Ht} \partial^2 \phi/\partial x^2+V'(\phi)=0
\end{equation}

The scalar field exhibits oscillon behavior for some forms of the nonlinear potential
$V$  (Fig. \ref{figosc}a), but not for others (Fig. \ref{figosc}b).

Where oscillons exist, a crude form of synchronized chaos is observed for
a pair of loosely coupled scalar field systems (a configuration that is
introduced to study the synchronization patterns, without physical motivation),
as seen in Fig. \ref{figoscsync}.  The fields do not synchronize, but the
oscillons in the two systems tend to form in the same locations.\footnote{The 
phases
of the oscillons do not necessarily agree, so  neither do we have an example
of {\it phase synchronization}  - the celebrated phenomenon \cite{Rosenblum,
Pikovsky} in
which a system that is chaotic can nevertheless be assigned a phase which
matches that of a second system. Oscillon frequencies depend on their
amplitudes, which are generally different for a pair of oscillons whose positions
correspond, and so the phases cannot agree. (Additionally, it is not clear how one would define
a phase for a multi-oscillon system that would capture information about their
positions.)} For a potential
that does not support oscillons, the positional coincidence is trivially
absent, and there is no correlation between corresponding components of the
underlying field.  Synchronization in this case can {\it only} be interpreted
in terms of coherent structures in the separate systems.

In a system as simple as the 3-variable Lorenz model, the hypothesis about
the relationship between internal and external synchronization is also
validated. In this case the relationship gives insight about which variables
can be coupled to give synchronized chaos.  Along the Lorenz attractor, the
variables $X$ and $Y$ partially synchronize, resulting in the near-planar
shape, while $Z$ is independent.  Consistently with the internal-external 
synchronization hypothesis, either $X$ or $Y$, but not $Z$, can be coupled to the
corresponding variable in an external system to cause the two systems to
synchronize, as is well known.

To summarize: The meaningfulness of a
synchronization pattern, as philosophically required, is naturally defined in terms of internal
synchronization, or coherent structures, involving some of the variables
that synchronize externally.  But external synchronization usually (or, by
hypothesis, always) implies
the existence of internal synchronization, and hence meaning.

\section{Sync as an Organizational Principle in Mind and in Computational Modeling}
\label{sec:comp}
If chaos synchronization provides a rational foundation for philosophical
synchronicity, it should give deeper insight regarding apparent synchronicity in physical
and psychological phenomena and underlying mechanisms.  In the psychological
realm, it has already begun to appear that synchronized oscillations play a key
role.  Synchronized firing of neurons has been introduced as a mechanism for
grouping of different features belonging to the same physical object
\cite{von86},\cite{Gray},\cite{Sch}. Debates over the physiological basis of consciousness
have centered on the question of what groups or categories of neurons must
fire in synchrony in a mental process for that process to be a ``conscious" one
\cite{Koch}.   It was argued previously that patterns of synchronized firing
of neurons provide a particularly natural and useful representation of objective
grouping relationships, with chaotic intermittency allowing the system to
escape locally optimal patterns in favor of global ones \cite{DuaChaos},
following an early suggestion of Freeman's \cite{Fre}.
The observed, highly intermittent synchronization of 40Hz neural spike trains
could play just such
a role.

The role of spike train synchronization in perceptual grouping has led to
speculations about
the role of synchronization in consciousness \cite{von86},\cite{Rod},\cite{Str},
\cite{Koch},
but here we suggest a
relationship on a more naive basis: 
The hallmark of conscious thought, defined subjectively, is the 
ability to focus on one's own thoughts, and to influence them, as one would 
interact with external events.  Thus consciousness can be framed as 
self-perception, and then 
placed on a similar footing as perception of the objective world.  In this
view, there must be
semi-autonomous parts of a ``conscious" mind that perceive one
another. In the interpretation of Section \ref{sec:percept}, these components
of the mind synchronize with one another, or in alternative language,
they perform ``data assimilation" from one another, with a limited
exchange of information.  The scheme has actually been proposed, and is 
currently
being investigated, for fusion of alternative computational models of the same
objective process in a practical context \cite{van,Mir}.

Taking the proposed interpretation of consciousness seriously, again imagine that
the world is a 3-variable Lorenz system, perceived by three different
components of mind, also represented by Lorenz systems, but with different
parameters. The three Lorenz systems also ``self-perceive" each other.
Three imperfect ``model"
Lorenz systems were generated by perturbing parameters in the differential
equations for a given ``real"
Lorenz system and adding extra terms.  The resulting suite is:
$\dot x = \sigma (y-z), \;\;
\dot y = \rho x - y - x z,\;\;
\dot z = -\beta z + x y$ 
\begin{eqnarray}
\label{adaptLor}
\dot x_i &=& \sigma_i (y_i - z_i) +\sum_{j\ne i} C^x_{ij} (x_j - x_i) 
+ K_x(x-x_i) \nonumber \\
\dot y_i &=& \rho x_i - y_i - x_i z_i + \mu_i +\sum_{j\ne i} C^y_{ij} (y_j - y_i)
+ K_y(y-y_i)  \\
\dot z_i &=& -\beta_i z_i + x_i y_i +\sum_{j\ne i} C^z_{ij} (z_j - z_i)+ K_z(z-z_i) \nonumber 
\end{eqnarray}
where $(x,y,z)$ is the real Lorenz system and $(x_i,y_i,z_i)\; i=1,2,3 $ are the three
models.  An extra term $\mu$ is present in the models but not in the real
system. Because of the relatively small number of variables available in
this toy system, all possible directional couplings among
corresponding variables in the three Lorenz systems were considered, giving
18 connection coefficients $C^A_{ij} \; \; \; A=x,y,z \; \; \;i,j= 1,2,3 \;\;\;
i\ne j$. 
The constants $K_A \;\;\;A=x,y,z$ are chosen arbitrarily so as to effect ``data
assimilation" from the ``real" Lorenz system into the three coupled ``model"
systems. The configuration is schematized in Fig. \ref{figsuite}.

The connections $C_{ij}$ 
linking the three model systems can be chosen using yet a further extension of 
the synchronization paradigm:  If two systems synchronize when their parameters 
match, then under some weak assumptions, as was proved in \cite{Dua07}, it
is possible to prescribe a dynamical evolution law for general parameters in
one of the systems so that the parameters of the two systems, as well as
the states, will converge. In the present case the tunable parameters are taken
to be the connection coefficients (not the parameters of the separate Lorenz
systems), and they are tuned under the peculiar assumption that reality itself 
is a similar suite
of connected Lorenz systems. The general result \cite{Dua07}
gives the following adaptation rule for the
couplings:
\begin{equation}
\label{adaptC}
\dot C^x_{i,j}= a (x_j - x_i) \left(x - \frac 1 3 \sum_k x_k \right) 
          - \epsilon/(C^x_{i,j} - C_{\rm max})^2  + \epsilon/(C^x_{i,j} + \delta)^2 
\end{equation}
with analogous equations for $\dot C^y_{i,j}$ and $\dot C^z_{i,j}$,
where the adaptation rate $a$ is an arbitrary constant and the terms with
coefficient $\epsilon$ dynamically constrain all couplings
$C^A_{i,j}$ to remain in the range $(-\delta,100)$ for some small number
$\delta$.  Without recourse to the formal result on parameter adaptation,
the rule (\ref{adaptC}) has a simple interpretation: Time integrals of the
first terms on the right hand side of each equation give correlations between
truth-model synchronization error, $x - \frac 1 3 \sum_k x_k $, and
inter-model ``nudging", $x_j - x_i$. We indeed want to increase or decrease
the inter-model nudging, for a given pair of corresponding variables, depending
on the sign and magnitude of this correlation. (The learning algorithm we have described resembles a supervised
version of Hebbian learning.  In that scheme ``cells that fire together wire
together."  Here, corresponding model components ``wire together" in a preferred
direction, until they ``fire" in concert with reality.) The procedure will 
produce a set
of values for the connection coefficients that is at least locally optimal in 
the multidimensional space 
of connection values.

A simple case is one in which each of the three model systems contains
the ``correct" equation for only one of the three variables, and ``incorrect"
equations for the other two. The ``real" system could then be formed 
approximately using large connections for the three correct equations, with other
connections vanishing, and the peculiar assumption is strictly true if the 
large connections become infinite.  Other combinations of model equations will also
approximate reality.
  
Several things have 
been learned from ``supermodels" such as the one defined by (\ref{adaptLor}) 
and (\ref{adaptC}). First, 
it is not difficult to define adequate inter-model connections.  In a numerical 
experiment (Fig. \ref{fig4Lor}a), the couplings
did not converge, but the coupled suite of ``models" did indeed synchronize
with the ``real" system, even with the adaptation process turned off half-way
through the simulation so that the coupling coefficients $C^A_{i,j}$
subsequently held fixed
values. Second,  the inter-model connections are needed, despite efforts, common in the 
modeling community \cite{TebKnutti}, to combine only the outputs of independently run 
models using Bayesian reasoning.  The difference between corresponding
variables in the ``real" and coupled ``model" systems was significantly less than
the difference using the average outputs of the same suite of models, not
coupled among themselves. (The three models also
synchronized among themselves nearly identically.) Further, without the model-model coupling, the output of the single model with the 
best equation for the given
variable (in this case $z$, modeled best by system $1$) differed even more from
``reality" than the average output of the three models. Therefore, it is
unlikely that any {\it ex post facto} weighting scheme applied to the three
outputs would give results equalling those of the
synchronized suite. Internal synchronization within the multi-model ``mind"
is essential. Third, the choice of 
semi-autonomous models to be combined is not essential, as long as the ``gene 
pool" of models is diverse.  In a case where no model had the ``correct" 
equation for any
variable, results deteriorated only slightly (Fig. \ref{fig4Lor}d).

The above scheme for fusion of imperfect computational/mental models only 
requires that the models come equipped with a procedure to assimilate 
new measurements from
an objective process in real time, and hence from one another. The scheme has 
indeed been proposed for the
combination of long-range climate projection models, which differ significantly
among themselves in regard to the magnitude and regional characteristics of expected global
warming \cite{DuaEGU}. (To project 21st century climate, the models are 
disconnected from reality after training, parameters are altered slightly to
represent increased greenhouse gas levels, and one assesses changes in the
overall shape of the attractor.) In this context, the previous results with Lorenz
systems were 
thoroughly confirmed and extended using a learning method that minimizes
synchronization error over finite-length trajectories, instead of the
instantaneous error as above,
to determine inter-model
connections \cite{van,Mir}. The scheme could also be applied to financial,
physiological, or ecological models.

In the realm of mind, the sharpening of the 
transition to synchronization as the suite of interconnected systems increases 
in size is taken here to bolster
the previous suggestions that synchronization plays
a fundamental role in conscious mental processing. (We have focussed on
assimilation/perception, but analogous constructions could be applied to the 
opposite problem of control - the interaction between mind and matter is
two-way.)   For application to mind, we imagine that the
systems are neurons or collections of neurons.  Note that the proposed role of
synchronization is markedly different from Pauli's view of mental
phenomena as ``something objectively psychical which cannot and should
not be explained by material causes"  \cite{HA, Meyenne}.  Here mental
phenomena are grounded in the material reality of neuronal systems, even if
their dynamical properties are qualitatively different from those of the much 
higher dimensional physical world that they represent.

To describe ordinary mental phenomena, one needs a notion of synchronization at
slower time scales and higher levels of organization, so that alternative representations of
the same objective reality within the brain can fuse to form 
a conscious percept.
Thus the synchronization 
view suggests a new direction of research, since it remains to integrate a 
theory of higher-level synchronization with the known
synchronization of 40Hz spike trains. It is certainly plausible that 
synchronization at higher levels could 
rest on synchronization at shorter time scales. Inter-scale interactions played a similar
role in the synchronization of a range of Fourier components of the same
field in the synchronously coupled systems of partial differential equations
considered in Section \ref{sec:percept}.  Conversely, feedback from the 
interpretive stages 
might trigger or reinforce the low-level synchronization.  This indeed must 
occur to explain the known neural phenomena in visual grouping \cite{Gray}, but 
realistic models remain to be constructed.   In such models, with a steady 
stream of new input data, natural noise or chaos would cause the periods of 
high-quality synchronization across the system to be brief (as in 
Fig. \ref{figsyncdelay}c) (cf. \cite{Ash}).  Analogous neural 
synchronicities at multiple levels of a processing hierarchy in a real 
organism would appear subjectively as consciousness.

\section{Sync in Quantum Theory}
\label{sec:quantum}
It has been asserted that an ``acausal connecting principle" applies not only to the
relationship between mind and matter, or to relationships within mind, but also
applies to matter itself.
Turning to the realm of basic physics, the fundamental role of synchronism is most 
evident in the surprising long-distance correlations that characterize
quantum phenomena. The Einstein-Podolsky-Rosen (EPR) phenomenon, viewed in
light of Bell's Theorem, implies that spatially separated physical 
systems with a common history continue to evolve as though connected with
each other and with observers.  Bell's Theorem definitively asserts that observed
correlations between two spatially separated spin-1/2 particles arising from the decay of
a common spin-0 ancestor cannot be explained in terms of a {\it causal} relationship 
to the initial decay conditions alone. Such a relationship would imply that the
binary-valued 
spins $A$ and $B$ are only functions of the orientations $a$ and
$b$ of the respective measuring devices, and of some hidden variables
represented collectively by $\lambda$, i.e. $A=A(a,\lambda)$, $B=B(b,\lambda)$.
In general, $\lambda$ designates the state of the joint system at some initial time.  One then 
defines the correlation $P$ between the two measured spins as a
function of the two orientations $a$ and $b$, 
$P(a,b)\equiv\int\rho(\lambda)A(a,\lambda)B(b,\lambda)d\lambda$,
where $\rho(\lambda)$ is any function specifying a probability distribution of the hidden variables, 
i.e. $\int\rho(\lambda)d\lambda=1$. But no matter how $\lambda$ is defined or how its values are
distributed, if $\lambda$ and $\rho(\lambda)$ are independent of $a$ and
$b$ the correlations $P$ are easily shown to satisfy a relationship that disagrees with standard
quantum theory and with experiment \cite{Bell}, negating the assumed causal relationship. 
The picture of the quantum world that emerges 
from Bell's Theorem is one in which entanglement is pervasive, even among
virtual particles in the vacuum, defining a web of relationships similar to the one
implied by the ubiquity of chaos synchronization.  As with synchronized chaos, quantum 
entanglement can be used for cryptography, an analogy that was developed
in prior work \cite{DuaFPL}.

One may 
naturally seek an interpretation of EPR correlations in terms of synchronized
chaotic oscillators, if one puts quantum theory on a non-local deterministic 
footing,
as first done in Bohm's interpretation \cite{Bohm52},\cite{Bohm93}. Bell's Theorem denies 
the possibility of either a quantum theory or 
an underlying hidden-variable theory which  is local, so some form of 
non-locality is required, but that non-locality should still not allow faster-than-light signals which would violate relativity.  Bohm, following de Broglie,
introduced an explicitly non-local potential to construct a theory that was entirely 
equivalent in its predictions to standard non-relativistic quantum theory. In a
two-slit experiment for example, a particle can follow one of two paths defined
by potential valleys, with the choice of path sensitively dependent on initial
conditions, as in a chaotic system.  
But the many-body form of the potential is typically  unanalyzable, 
an issue that arises especially in extension to relativistic 
field theory.  The view taken here is that Bohm's interpretation is an
``existence proof" of the possibility of a hidden variable theory, countering 
von Neumann's earlier alleged proof that such theories are impossible
\cite{vonNeumann}, and not a satisfactory theory as it stands.

In 't Hooft's more rece
interpretation \cite{tHooft99,tHooft09}, which might 
ultimately generate testable predictions at the microscale, there is a weaker 
form of non-locality due to the presence of Planck-scale black holes, and an
essential entanglement between the choices made by an observer ($a$ and $b$ 
above) and the observed state, as must arise since ``free will" in making such 
choices is incompatible with determinism.  New conservation principles, not 
articulated as yet, are posited to constrain both physical states and observer 
choices \cite{tHooft09}.  In this Section,  in accordance with Bohm's  
coordinated ``ballet dances" of particles\cite{Bohmdance}, and with
 't Hooft's  new conservation principles, we extend a previous  
 speculation \cite{Dua05} that chaos synchronization can contribute
to a realist interpretation of quantum theory.

In an interpretation based on a granular state-space as 
with 't Hooft's, Palmer \cite{Pal09}
has hypothesized that the quantum world lives on a dynamically
invariant fractal point set within the higher-dimensional phase space 
associated with the degrees of freedom that are naively thought to be
independent. Membership in the invariant set is an uncomputable property,
so theories can only be formulated in terms of the variables of the full
phase space, and the emergent  
``conservation laws"  that restrict motion to the invariant set remain implicit.
Palmer's invariant set is in fact a generalized synchronization
``manifold" (the common but improper term, since the ``manifold" is nowhere
smooth), of the sort suggested by Fig. \ref{figgensync}c. As discussed in prior
work \cite{Dua05},
generalized rather than identical
synchronization is the fundamental relationship because EPR spins {\it
anti-}correlate.

To bar supraluminal transmission of information, we rely on the mechanism
proposed for synchronization-based cryptography: a signal provided by one
variable of a chaotic system is difficult to distinguish from noise and
is meaningful only when received by an identical copy of the system.  However,
it follows from Takens theorem \cite{Takens} that information can be extracted from such
a signal if one considers a long enough time series. Longer time series
are required to decode signals produced by more complex systems. For perfect
security, one would need a chaotic system with an infinite-dimensional
attractor. 

Such a situation would arise most naturally in a multi-scale system (e.g.
as proposed by Palmer\cite{Pal04})
requiring at least a system of partial differential equations, but something can be 
learned by considering a family of simpler systems of variable dimension, given
by ordinary differential equations \cite{DuaFPL},\cite{Dua05}. 
It is known that two $N$-dimensional Generalized Rossler
systems (GRS's) (each equivalent to a Rossler system for $N=3$) will synchronize for any
$N$, no matter how large, when coupled via only one of the $N$ variables:

\begin{minipage}[t]{3.5in}
\begin{eqnarray*}
\hspace{.8in}\dot{x}^A_1&=&-x^A_2 + \alpha x^A_1 + x^B_1 - x^A_1 \nonumber \\
\dot{x}^A_i&=&x^A_{i-1} - x^A_{i+1} \nonumber \\
\dot{x}^A_N&=&\epsilon + \beta x^A_N (x^A_{N-1} - d) \nonumber\\
\end{eqnarray*}
\end{minipage}
\begin{minipage}[t]{3in}
\begin{eqnarray}
\label{cGRS}
\dot{x}^B_1&=&-x^B_2 + \alpha x^B_1 + x^A_1 - x^B_1 \nonumber \\
\dot{x}^B_i&=&x^B_{i-1} - x^B_{i+1}  \hspace{.25in} i=2\ldots{N-1} \hspace{.5in}\\
\dot{x}^B_N&=&\epsilon + \beta x^B_N (x^B_{N-1} - d) \nonumber
\end{eqnarray}
\end{minipage}\\
Each system has an attractor of dimension $\approx N-1$, for $N$ greater than
about 40, and a large number of positive Lyapunov exponents that increases
with $N$. As $N\rightarrow\infty$, while the synchronization persists, the
signal linking the two systems becomes impossible to distinguish from noise.
It was shown previously \cite{DuaFPL} that an inequality analogous to Bell's could
be constructed by arbitrarily bisecting the phase space to define final states
analogous to spin-up/spin-down, and using a GRS parameter as an analogue of
measurement orientation. That inequality is in fact
violated because of the connection between the systems, but a naive observer
would expect it to hold because he is unable to distinguish the connecting
signal from noise.

In Palmer's view, there is no connecting signal because the world never leaves
the ``invariant set" (although the dissipative character of
gravitational interactions is assumed to play a role cosmologically in
dynamically constraining the universe to motion on the invariant set in the first
place) \cite{Pal09}.  Here we consider the nature of the required ``restoring force" if
small perturbations transverse to the synchronization manifold are conceived
as physical. 

The GRS is a questionable model of reality because its largest
Lyapunov exponent $h_{max}\rightarrow0$ as $N\rightarrow\infty$  (the system's
``metric entropy," the sum of the positive Lyapunov exponents, 
$\sum_{h_i>0} h_i$, is constant as $N\rightarrow\infty$). In other
words, the higher the dimension, the less chaotic the system. Such behavior
is suspect in a system intended to represent unpredictable quantum fluctuations.  
Taking the GRS behavior as $N\rightarrow\infty$ to be generic, one must
reconcile its increasingly mild character with the requirement that the 
nonlocal ``signal" be perfectly masked through chaos.  It was noted
previously
\cite{Dua05} that the issue is resolved if 
the GRS is viewed as a spatially asymptotic description 
of an intrinsically faster dynamics in a highly curved space-time. For
reference, recall that an object falling into a black hole is perceived by
an observer at a distance from the hole as approaching the horizon with
decreasing velocity, but never reaching it. 
If the physical system that the GRS describes lives in the vicinity of a
micro- black hole or wormhole, the variables in the asymptotic description will
be slowed, but the actual physical processes will be realistically violent, and
can couple to each other through ``signals" that are perfectly masked. More 
generally, the synchronizing subsystems can be expected 
to behave more wildly than the usual systems defined by  PDE's on a continuum 
(cf. [40]) .  Some form of granularity in state-space  and/or physical 
space-time is indicated, in agreement with the models of 't Hooft 
\cite{tHooft09} and of 
Palmer \cite{Pal04,Pal09}.

\subsection{Physical vs. Virtual Non-locality}

A Planck-scale foam-like structure in space-time was posited by Hawking
\cite{Haw78}
in the context of a procedure to quantize classical general
relativity where that structure
contributes significantly to a sum over alternative {\it Euclidean} space-time 
geometries. The question here is about the possible role of
microwormholes in long-range synchronization, without transmission of
information, in ordinary Lorentzian space-time. That
possibility is consistent with theoretical arguments \cite{Bom} and 
experimental evidence \cite{AC} for fundamental granularity in
space-time structure. In the Appendix, it is explained that microwormholes could 
arise
in a variant of general relativity defined by equations that are generally 
covariant but scale-dependent, and a weak divergence that arises from the
recirculation of virtual quanta through wormholes may be avoided if
the wormholes are sufficiently narrow. Our wormholes
are reminiscent of those in the original construction of Wheeler \cite{Whe},
who suggested that lines of electric force are always closed if positive
and negative charges are thus connected at the microlevel. In the absence of
a theory of Planck-scale processes, other forms for the non-local physical 
connection could be considered, but general-relativistic wormholes are a 
natural starting point for any such construction.  Indeed, Maldacena and
Susskind have recently suggested that non-traversible wormholes mediate EPR
correlations in standard quantum theory \cite{MS2013}. A modicum of
traversibility, as described in the Appendix, could then support a non-local
deterministic interpretation.

If connections are formed by joining micro black holes, as might coincide with 
particles in an EPR pair, the systems that must 
synchronize are defined on two-dimensional horizons
at the mouths of the wormholes. It is consistent with the holographic principle
\cite{tHooft93},\cite{Sus} that such 2D fields capture the essential information about
the full three-dimensional systems. 

On the other hand, if we stipulate, with Palmer, that the synchronization manifold is fundamental,
because the physical world never leaves it, then no wormholes are needed:
We have two dynamical systems defining an anticorrelated EPR pair, 
\newcommand{\bx}{{\bf x}}
\newcommand{\by}{{\bf y}}
$\dot{\bx} = F(\bx)$, and $\dot{\by} = G(\by)$,with $\bx\in R^N$ and 
$\by\in R^N$. The dynamics are modified so as to
couple the systems:
\begin{equation}
\label{eqdscoup}
\dot{\bx} = \hat{F}(\bx,\by) \hspace{1in}
\dot{\by} = \hat{G}(\by,\bx)
\end{equation}
and there is some locally invertible function 
$\Phi:R^N\rightarrow R^N$ such that
$||\Phi(\bx)-\by||\rightarrow 0$ as $t\rightarrow\infty$. Then the coupled
dynamics are also defined by the two autonomous systems
\begin{equation}
\label{eqdscoup1}
\dot{\bx} = \hat{F}(\bx,\Phi(\bx)) \hspace{1in}
\dot{\by} = \hat{G}(\by,\Phi^{-1}(\by))
\end{equation}
without recourse to wormholes or any nonlocal connections, provided we know the badly behaved
function $\Phi$ exactly.
Otherwise, we rely on the narrow width of the wormholes to prevent supraluminal transmission
of matter or information.  Diffraction effects preclude communication, except
in highly symmetrical situations, as in EPR, where constructive interference
would account for the needed nonlocal connections. The isolated
character of such quantum ``synchronicities" follows from the rarity of the
required symmetrical context. 

The existence of the wormholes could shape the
synchronization manifold and the correspondence $\Phi$,  whether or not the
connections continue to play a role in the dynamics. That sparse 
connections can be sufficient to synchronize two
extended systems
has already been demonstrated.
Kocarev {\it et al.} \cite{Koc} showed
that pairs of PDE systems of various types (Kuramoto-Sivishinsky, complex
Ginsburg-Landau, etc.) could be synchronized by pinning corresponding variables 
to one another at a discrete set of points, at discrete instants of time.
(The example of synchronizing two quasigeostrophic channel
models (Fig. \ref{figQGsync}) establishes essentially the same phenomenon for
coupling formulated in Fourier space.)  A sparse set of wormholes is expected to
give synchronization of  subsystems on opposite sides in the same way. Once
the subsystems are perfectly synchronized, physical connections are no longer
needed, if we can rationalize the continued relevance of Eq. \ref{eqdscoup1}.

The mediation of quantum interconnectedness  by wormholes, temporary or lasting, is
perhaps the ultimate home for the marriage \cite{Str} between synchronization dynamics
and small-world (or ``scale-free") networks. The proposal would
also realize the program, favored by a minority of physicists, of quantizing gravity by rooting quantum behavior in 
space-time geometry,
rather than the reverse.  The question
is essentially whether the construction can reproduce the nonlocal piece
of the ``quantum potential" in Bohm's interpretation \cite{Bohm52} (the
remaining piece corresponding to motion along the synchronization manifold),
accounting for the origin of that piece geometrically.

In the 
synchronization framework, one could also imagine classical entanglement between 
observer choices and observed states, again with no explicit description of
the relationship.  The degrees of freedom corresponsding to observer choice must
in fact be part of the definition of the overall synchronization manifold on
which the world resides.  The synchronized chaos 
framework thus supports a local contextual resolution of the Bell paradox, as
well as a
natural description of  the weak nonlocality that is required if observer free
will is assumed.  Within the single framework, non-local connections 
 may or may 
 not be present accordingly, or might be vanishingly small and conceivably rooted in 
 cosmological history.

\section{Summary and Concluding Remarks}
\label{sec:concl}
In the foregoing sections, we have attempted to show that the synchronization
of loosely coupled chaotic systems approaches the philosophical notion of
highly intermittent, meaningful synchronicity more closely than commonly
thought.  Synchronized chaos is highly intermittent in a natural setting
(Section \ref{sec:back}). As with philosophical synchronicity, it 
describes the relationship between the objective world and a perceiving mind
or computational model (Section \ref{sec:percept}). The phenomenon typically involves the
coincidence of coherent structures, to which meaning can be attached (Section
\ref{sec:meaning}). Central to our thesis is a relationship between internal
synchronization within a system, and external synchronizability with another
physical system or with a model. That relationship, which was described
in Section \ref{sec:meaning}, is in accord with
common wisdom: An objective system with a high degree of internal
synchronization is more easily perceived/understood, an internally coherent 
individual can more easily engage the
world,  etc.  And accordingly, that relationship is currently proving to be a
useful design principle in computational modeling, as described in Section
\ref{sec:comp}.

What is not clear is that even with the isolated character and meaningfulness
of synchronicities in coupled chaotic systems, the phenomenon reaches all the way
to that of Jung and others, who discussed detailed coincidences between
physical events and previous dreams, for instance. The attempt to put
relationships of that kind on a rational footing may appear doomed. The
mechanisms of deterministic chaos seem  barely sufficient.  One may dismiss such
examples and
consider only more restricted forms of synchronicity, or one may imagine
that new physical principles emerge. The difficulty of ascribing the more
extended notion of synchronicity to material reality may indeed have led to the
dual-aspect monist conjecture \cite{HA} in which mind is elevated to the same level as
matter, and both are aspects of an underlying domain that is neither mental
nor material.  In contrast, although mind is semi-autonomous in the picture
presented here, our view is decidedly materialist.
Our endeavor might be compared to Marx's attempt
to ground Hegel's dialectic in material reality, a transformation whose
legitimacy has sometimes been questioned, notably by Bohm \cite{Peat96}.  

Closer to the physics is the example 
of Einstein's relationship to the ideas of Ernst Mach.  Einstein was inspired by 
Mach's relativism, but Mach castigated Einstein for the latter's belief 
that atoms are real, preferring to view them only as useful conceptual 
constructs \cite{Machbio}.  As science progresses, it is to be expected that
some ideas
previously introduced on a religious basis, or by idealist philosophers, are not 
wrong, but have indeed a hidden justification in material reality.  Even if our
objective realization of the Jung-Pauli notion is only partial, it is the point of view of
this paper that it is appropriate for scientists to seriously consider a concept
that has captured the popular imagination as widely as has syncrhonicity, and to
afford a rational explanation if possible. So the 
question is whether nonlinear dynamics has gone far enough as to put 
philosophical synchronicity on an objective footing to any significant degree.

The question is perhaps sharpest in regard to consciousness and
synchronization-based theories thereof. In Section \ref{sec:comp}, it was
argued that previous suggestions about the role of synchronization in the brain
were supported by the possibility of highly intermittent synchronization
among chaotic oscillators and by the possibility of synchronizing different
complex models of the same objective process, giving rise to ``self-perception".
But Penrose has given a  well known argument  that the reasoning abilities of conscious
beings cannot arise from classical physics or algorithmic processes that
describe such physics:    For any algorithmic system of ascertaining truth,                   
one can always articulate a true statement, of the sort constructed by
G\^odel, that such a being knows to be true,
but whose truth cannot be established within the system \cite{Pen}.
Since synchronized chaos is still deterministic\footnote{If one considers
a chaotic system given by differential equations for which infinite precision
in initial conditions is needed to predict the outcome even qualitatively, as in
Palmer's earliest proposal \cite{Pal95}, a typical basin of attraction for a given
outcome is a ``fat fractal": The more precisely the initial conditions are
known,  the smaller is the probability of error in ``guessing" the outcome. That
is very unlike quantum indeterminacy.}, the abilities of conscious
beings must come from fundamentally different processes, which Penrose has
concluded are quantum mechanical.

The discussion of quantum processes in Section \ref{sec:quantum} was included
because they
seem to provide the deepest example of synchronicity - the quantum world
appears to live on a generalized synchronization ``manifold". But if Penrose
is correct, the converse statement can also be made: Synchronicity as
manifest in human consciousness is also fundamentally quantum in origin.
Correlations in neuronal firing or between neural subsystems can only give
rise to consciousness, in this view, if quantum correlations are 
involved, such correlations arising either as in standard quantum theory or
as in deterministic re-interpretations thereof. Synchronicities between states of the mind and of the objective
world must somehow follow. Perhaps such an enlarged notion could reach the
popular concept, and the one of Jung and Pauli. In Penrose's view, as here,
the question of the proper interpretation of quantum phenomena on the
one hand, and that of the origin of synchronism between mind and matter on
the other, are to be resolved jointly. Our proposal differs in that dynamical
synchronization, in a properly structured microworld, would account for quantum
correlations, but would also explain macroscopic phenomena, mental and
material, directly.

{\bf Acknowledgements:} The author is grateful for discussions over the past
decade with Joe Tribbia, Jeff Weiss, Alan Guth, Frank Hoppensteadt, Ljupco
Kocarev, Tim Palmer, and David Peat. Part of this work was supported under 
NSF Grants \#0327929 and \#0838235, DOE Grant \#DE-SC0005238, and
European Research Council Grant \#266722.

\appendix
\section{On the Possibility of Microwormholes}

As discussed in Section \ref{sec:quantum} of the text, if space-time were permeated with 
micro-scale Wheelerian wormholes, that would be useless for time travel, a synchronistic
order could nonetheless emerge: chaos synchronization could combine with small-world
effects in the same manner as has been described for more familiar
applications. Consideration of synchronicity as a basic physical principle
would not be complete without considering Wheeler's suggestion.  In a deterministic theory 
underlying quantum mechanics, a wormhole might connect the charges in an EPR 
pair. The role of wormholes in mediating quantum entanglement posited by
Maldacena and Susskind \cite{MS2013} would be enlarged in a non-local
deterministic theory. Here we discuss two historical objections to wormholes 
as they would impact their possible occurrence at the microscale.\\

\noindent{\bf A.1 Implications of the weak-energy condition in ordinary and higher-derivative
gravity}\\

While two Schwarzchild black hole solutions to Einstein's
equations can be joined to form a wormhole, solutions of this type are
not traversible \cite{MorAJP}. The possibility of traversible wormhole solutions is
limited by the {\it weak energy condition}. That condition states that for
any null geodesic, say one parameterized
by $\zeta$, with tangent vectors $k^a=dx^a/d\zeta$, an averaged energy
along the geodesic must be positive: 
\begin{equation}
\label{WEC}
\int_0^\infty T_{\alpha \beta}k^\alpha
k^\beta > 0
\end{equation}
where $T_{\alpha \beta}$ is the stress-energy tensor. Traversible wormholes 
can 
exist only if (\ref{WEC}) is violated for some null geodesics passing through
 the wormhole, implying the existence of ``exotic matter" with 
negative energy density in the ``rest frame" of a light beam described by
the null geodesic. The negative energy density is required, in one sense,
to hold the wormhole open.

Quantum fluctuations in the vaccuum can violate the weak energy condition
\cite{MorPRL},\cite{Can}. But the problem can be avoided at the
classical level, as desired if quantum theory is not to be presumed, if a larger class of generally covariant
theories are considered.  Terms containing higher derivatives of the metric can
indeed be added to Einstein's equations, with effects that are negligible on
all but the smallest scales \cite{Wei},\cite{Ste}. The situation is analogous to that of the Navier-Stokes
equation in fluid dynamics: While the terms involving the co-moving derivative follow
simply from Newton's first law, the dissipative terms are ad hoc and can take many forms.
General relativity can likewise be extended to theories of the form: 
\begin{equation}
\label{modGR}
R_{\mu \nu} -\frac{1}{2}Rg_{\mu \nu} + g_{\mu \nu}\Lambda 
+ \sum_{n>2} c_n L^{n-2} R^{(n)}_{\mu \nu} = 8\pi T_{\mu \nu}
\end{equation}
where  $R^{(n)}_{\mu \nu}$ is a quantity involving a total of $n$ derivatives of the
metric, $L$ is a fundamental length scale, the $c_n$ are dimensionless
constants, and we have included a cosmological constant $\Lambda$ for full generality. If $L=L_P$, the Planck length, then the new
terms in the extended theory (\ref{modGR}) are negligible on macroscopic scales. They only need be considered
if curvature is significant at the Planck length scale.  Any metric
that solves the ordinary Einstein equations after the substitution
$T_{\mu \nu} \rightarrow T_{\mu \nu} - (1/8\pi)\sum_{n>2} c_n L^{n-2}
R^{(n)}_{\mu \nu}$ solves (\ref{modGR}) for given $T_{\mu \nu}$.  It is plausible that 
the modified stress-energy tensor $T_{\mu \nu} - (1/8\pi)\sum_{n>2} c_n L^{n-2} R^{(n)}_{\mu \nu}$ can be made to violate the weak energy condition if
the signs of the constants $c_n$ are chosen appropriately, and thus that
a traversible micro-wormhole solution is possible.\\

\noindent{\bf A.2 Vaccuum recirculation effects for narrow wormholes}\\

The paradoxes that one normally associates with closed time-like curves, as 
would pass through
a wormhole, have a quantum counterpart: Repeated passage of a virtual particle through a 
wormhole may lead to a divergence in the stress-energy tensor $T_{\mu \nu}$.  The derivation of
this controversial result is as follows: For each 
passage of a virtual particle through the wormhole, the
contribution to the two-point function 
$<\Psi|\hat\phi(x)\hat\phi(x')+\hat\phi(x')\hat\phi(x)|\Psi>$
from a trajectory that contains
that passage is attenuated by a factor $b/D$, where $b$ is the wormhole
width, and $D$ is the spatial length of a geodesic through the wormhole,
as measured in the frame of an ``observer" traveling along the geodesic
from the vicinity of $x$ and $x'$ through the wormhole once and back
to the same vicinity.  Here, $x$ and $x'$ are nearby points in space-time,
$|\Psi>$ is the quantum state, and $\hat\phi$ is the field operator associated with the field $\phi$. The contribution to the two-point
function is found to behave as $~(b/D)^k \times 1/\sigma$,
where $\sigma$ is $1/2$ the square of the proper distance between $x$ and $x'$ 
along the geodesic connecting
them through the wormhole, and the power $k$ depends on the number of times
that the trajectory  traverses the wormhole. (Contributions from trajectories
that traverse the wormhole only once dominate.) One finds $\sigma \sim
D \Delta t$, where $\Delta t$ is the proper time between $x$ and the nearest
null geodesic that passes through the wormhole. As $x'\rightarrow x$,
the contribution diverges if $x$ can be joined to itself by a null geodesic
that passes through the wormhole.  The stress-energy tensor can be expressed
in terms of the two-point function \cite{Kim} and also
diverges as $\sigma\rightarrow 0$ or $\Delta t \rightarrow 0$. Specifically, one
finds $T_{\mu \nu}\sim(b/D)^k \times 1/D (\Delta t)^3$ in 
natural units, or in dimensional units,
\begin{equation}
\label{Tdim} 
T_{\mu \nu}\sim(b/D)^k \times L_P/D \times m_P/(\Delta t)^3.
\end{equation}


Kim and Thorne \cite{Kim} argued that the divergence, which is small because of the
``diffraction" factor $b/D$, probably disappears in the proper
quantum theory of gravity, allowing wormholes to remain. Quantization
of the gravitational field in that theory would be effective on scales of $L_P$, the Planck
length, so we only need consider the magnitude of $T_{\mu \nu}$ for
$\sigma \ge L_P$.  At these scales, referring to (\ref{Tdim}), 
$T_{\mu \nu} \le L_P/D$ in natural
units of $m_P/L_P^3$, giving energy densities that are far too weak to destroy
the wormhole, or have other noticeable effects, for macroscopic $D$.   

Hawking \cite{Haw92}, in support of his ``chrononology
protection conjecture", provided a counter-argument asserting that quantum
gravity effects would only enter on much smaller scales, corresponding to
the Planck length in the rest-frame of an ``observer" travelling on one
of the geodesics through the wormhole.  The values attained by $T_{\mu \nu}$
on scales larger than Hawking's reduced length scale would still cause collapse of the
wormhole, the instant that recirculation becomes possible.

Let us assume that the predictions of standard quantum theory in curved
spacetime survive in whatever deterministic theory underlies quantum
mechanics. Here, we note that there is an additional mechanism that might cut off the
recirculation divergence for wormholes of very narrow width.  Virtual
particles of arbitrarily high energy cannot traverse the wormhole. High-energy
virtual particles would reverse the effect of the exotic matter or of the
higher-derivative terms,  so the existence of the wormhole would again be
precluded by the weak energy condition. The contribution to the energy flux
from the virtual particles is  $T^{0i}=\frac{4}{\pi b^2}\int d\omega
n(\omega)\hbar \omega$, where $n(\omega)$ is the number
density of quanta at frequency $\omega$. (As in the Weizsacker
-Williams approximation \cite{Jac}, the quanta are assumed not to 
elongate in the wormhole.)  At detailed resolution in frequency-space, $n(\omega)=\sum_i n_i\delta(\omega-\omega_i)$, where ${\omega_i}$ is a
discrete set of frequencies and $\{n_i\}$ is a set of positive integers. There is a problem from the weak energy condition  if
any $\omega_i > \omega_{cutoff}$  (with $n_i\ge 1$), for
$\omega_{cutoff}$ sufficiently
large as to cancel the negative-energy contributions to  $T^{00}$.  In a path
integral, taken both over particle trajectories and over geometries, one need 
only consider histories in which  more energetic particles either
collapse the wormhole or are reflected and do not traverse it. In contrast, for wormholes
of macroscopic width, histories must be included in the path integral for which
the energies of recirculating virtual quanta outside the wormhole are
anomalously large (treating the geometry itself classically). The cutoff in
the former case implies that  
the term $1/\sigma$ in the two-point function is replaced by a term 
like $\int_{\omega_k <
\omega_{cutoff}} d^4 k \; \exp[i k\cdot (x-x')]/k^2$  which does not diverge. 

Thus for sufficiently narrow wormholes, the original position of Kim and Thorne that
that vaccuum recirculation divergence is damped may be correct. The behaviour
discussed above is on the borderline between the nontraversibility of
Einstein-Rosen bridges, which mediate entanglement
in ordinary quantum theory according to the recent proposal
\cite{MS2013}, and traversible wormhole behaviour that could give rise to
full nonlocality. Although finite wormhole lifetime would be required to mediate long-range 
synchronization, 
highly intermittent wormhole behavior may be enough, in accordance with previous 
findings \cite{Koc} for other types of PDE's.  And the collapse of a wormhole
triggered by the entrance of a particle on one side would have measurable
effects on the other side. In the absence of a detailed Planck-scale
theory, the question of traversibility remains open.


\newpage

\begin{figure}[H]
\caption{Diagram constructed by Carl Jung, later modified by Wolfgang Pauli,
to suggest relationships based on synchronicity as an ``acausal connecting
principle'', existing alongside causal relationships \cite{Jung}.}
\label{figJung}
\end{figure}

\begin{figure}
\caption{The trajectories of the synchronously coupled Lorenz systems 
in the Pecora-Carroll complete replacement scheme (\ref{eqPC}) rapidly
converge (a). Differences between corresponding variables approach
zero (b).}
\label{figPC}
\end{figure}

\begin{figure}[H]
  \caption{Transition from identical to generalized synchronization, illustrated 
 by the relationship between a pair 
 of corresponding variables $x$ and $x'$:  Projection of the synchronization manifold onto the $(x,x')$ 
 plane are shown for (a) identical synchronization, (b) generalized synchronization
 with near-identical correspondence, (c) generalized synchronization with a 
 correspondence function that is nowhere smooth.}
 \label{figgensync}
 \end{figure}

\begin{figure}[H]
   \caption{The difference between the simultaneous states of two Lorenz systems 
 with time-lagged coupling (\ref{PCdelay}), with $\sigma=10.$, $\rho=28.$, 
 and $\beta=8/3$, represented by 
 $Z(t)-Z_1(t)$ vs. t for various values of the inverse time lag $\Gamma$
 illustrating complete synchronization (a), intermittent or ``on-off''
 synchronization (b), partial synchronization (c), and de-coupled systems (d).
Average euclidean distance $\langle D \rangle$ between the states of the two systems in 
$X,Y,Z$-space is also shown. A histogram of the lengths of periods of ``synchronicity",
such as the one indicated by the arrow in (c), is shown in (e) for the time-delayed coupling
case (solid line) and a case of two unrelated Lorenz trajectories (dashed line),
 where synchronicity intervals are periods during which $|Z(t)-Z_1(t)|< 5$. }
  \label{figsyncdelay}
 \end{figure}  

\begin{figure}[H]
\caption{Streamfunction (in units of $1.48\times 10^9 m^2s^{-1}$) describing
the forcing $\psi^*$ (a,b), and the evolving flow $\psi$ (c-f), in a 
parallel
channel model with coupling of medium scale modes 
for which
$|k_x| > k_{x0}=3$ or $|k_y| > k_{y0}=2$, and $|k|\le 15$, for the
indicated numbers $n$ of time steps in a numerical integration (generalizing to
bidirectional coupling, for
convenience).  Parameters
are as described previously \cite{Dua04}. An average
streamfunction for the two vertical layers $i=1,2$ is shown. 
 Synchronization
occurs by the last time shown (e,f), despite differing initial conditions.}
\label{figQGsync}
\end{figure}

\begin{figure}[H]
\caption{Streamfunction (in units of $1.48\times 10^9 m^2s^{-1}$) describing 
a typical blocked flow state (a) and a typical zonal
flow state (b) in the two-layer quasigeostrophic channel model. 
 An average streamfunction 
for the two vertical layers 
$i=1,2$ is shown. }
\label{figblockzon}
\end{figure}

\begin{figure}[H]
\caption{Energy density 
$\rho = (1/2) e^{-Ht}(\phi_x)^2 + (1/2) e^{Ht}(\phi_t)^2 
+ e^{Ht} V(\phi)$,  vs. position $x$ for a numerical simulation of
the scalar field equation (\ref{osc}) with the potential $V(\phi)= (1/2) \phi^2 - (1/4) \phi^4
+ (1/6) \phi^6$, exhibiting localized oscillons (a), and a simulation
of the same equation, but with a different potential
$V(\phi)= (1/2) \phi^2 + (1/4) \phi^4
+ (1/6) \phi^6$, for which oscillons do not occur (b).}
\label{figosc}
\end{figure}

\begin{figure}[H]
\caption{The local energy density $\rho$ vs. x for two simulations of the scalar
field equation (\ref{osc}),
coupled to one another only through modes of wavenumber $k \le 64$,
where modes up to $k_{max}=2^{14}$ are realized numerically.  
($\rho$ for the second system (dashed line) is also shown reflected across the 
x-axis for ease in comparison.) The coincidence of oscillon positions is apparent.
}
\label{figoscsync}
\end{figure}

\begin{figure}[H]
\caption{``Model" Lorenz systems are linked to each other,
generally in both directions, and to ``reality" in one direction. Separate links
between models, with distinct values of the connection coefficients $C^{ij}_l$,
are introduced for different variables and for each direction of possible
influence.}
\label{figsuite}
\end{figure}

\begin{figure}[H]
\caption{Difference $z_m - z$ between ``model" and ``real" $z$ vs. time for a Lorenz
system  with $\rho=28$, $\beta=8/3$, $\sigma=10.0$ and an interconnected suite
of models with $\rho_{1,2,3}=\rho$, $\beta_1 = \beta$, 
$\sigma_1=15.0$, $\mu_1= 30.0$, $\beta_2=1.0$, $\sigma_2=\sigma$, 
$\mu_2 = -30.0$,
$\beta_3=4.0$,$\sigma_3=5.0$, $\mu_3=0$. The synchronization error is shown
for a) the average of the coupled suite $z_m=(z_1 + z_2 + z_3)/3$ with
couplings $C^A_{ij}$ adapted according to (\ref{adaptC}) for $0<t<500$ and
held constant for $500<t<1000$; b) the same average 
$z_m$ but with all $C^A_{ij}=0$; c) $z_m=z_1$, the output of the model with the
best $z$ equation, with $C^A_{ij}=0$;\; d) as in a) but with $\beta_1=7/3$, $\sigma_2=13.0$,
and $\mu_3=8.0$,  so that no equation in
any model is ``correct". (Analogous comparisons for $x$ and $y$
give similar conclusions.)}
\label{fig4Lor}
\end{figure}

\setcounter{figure}{0}

\protect\newpage
\begin{figure}
\vspace{5in}
\includegraphics{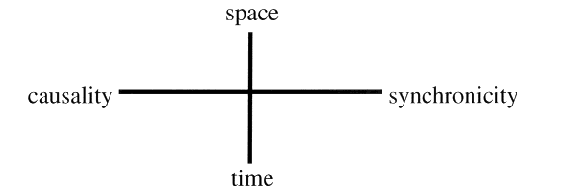}
\caption{}
\end{figure}

\newpage
\begin{figure}
\vspace{5in}
\begin{minipage}{5.5in}
   a) \resizebox{.5\textwidth}{!}{\includegraphics{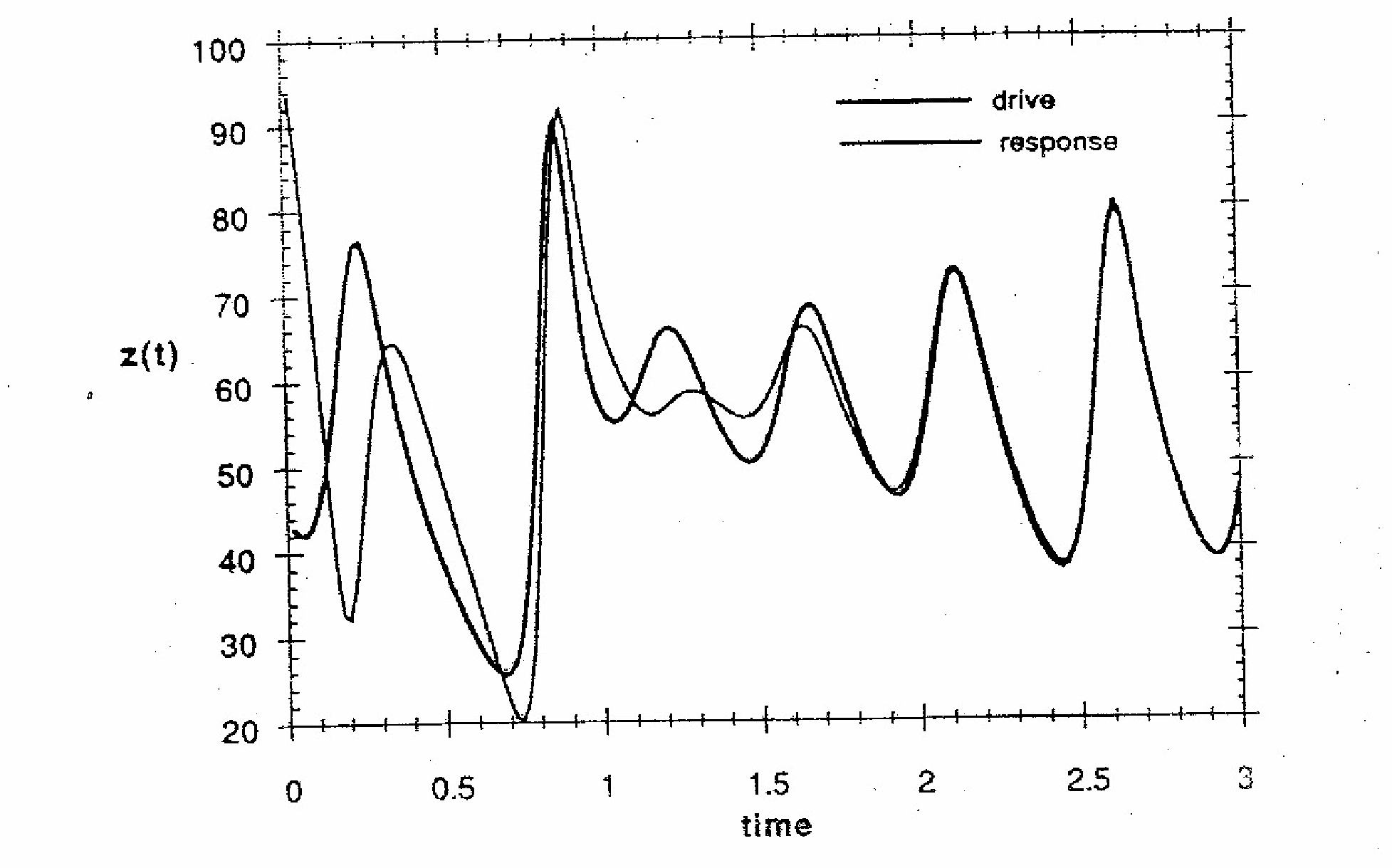}} \hspace{.4in} 
   b)\resizebox{.3\textwidth}{!}{\includegraphics{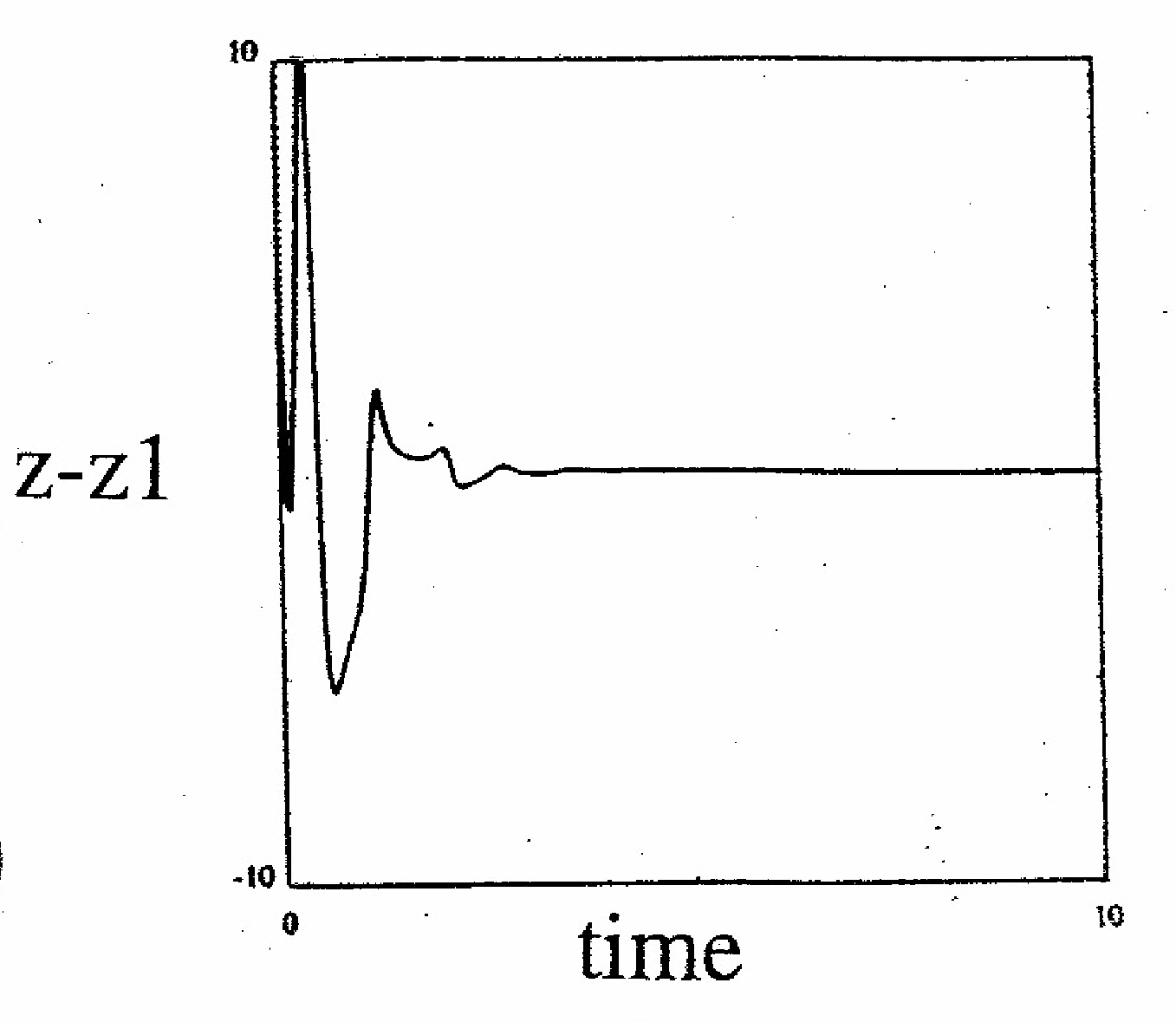}}
 \end{minipage} 
\caption{}
\end{figure}

\newpage
\begin{figure}
\vspace{4in}
   \begin{minipage}{7in}
   \vspace{.3in}
   a) \resizebox{.275\textwidth}{!}{\includegraphics{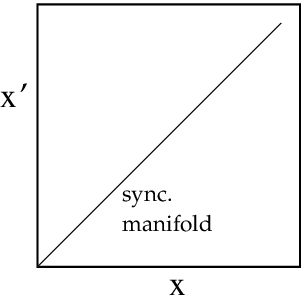}}  
   b) \resizebox{.275\textwidth}{!}{\includegraphics{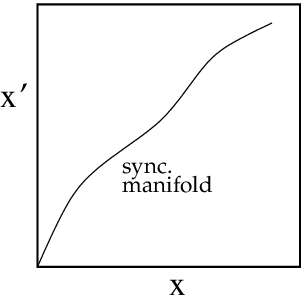}}
   c) \hspace{.05in}
   \resizebox{.275\textwidth}{!}{\includegraphics{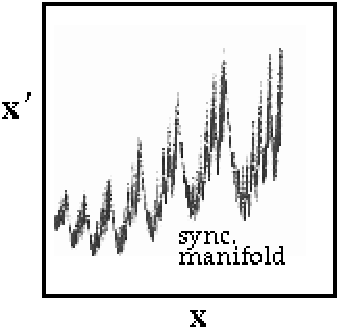}}
 \end{minipage}
  \vspace{1in}
   \caption{}
\end{figure}

\newpage
\begin{figure}
\begin{minipage}{5.5in}
   \begin{minipage}{7in}
   a) \hspace{.025in} \includegraphics{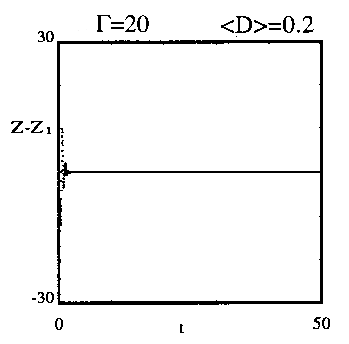} \hspace{.4in} 
                          b)\includegraphics{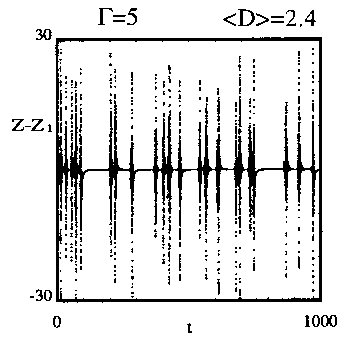}
 \end{minipage}
   \begin{minipage}{7in}
   \vspace{.3in}
  c) \includegraphics{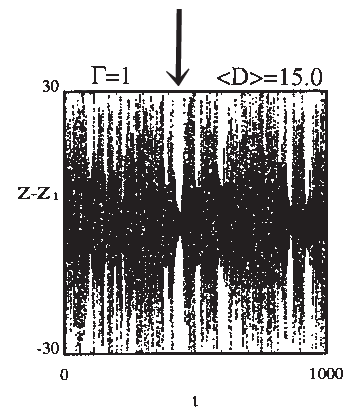} \hspace{.36in} 
        d)\includegraphics{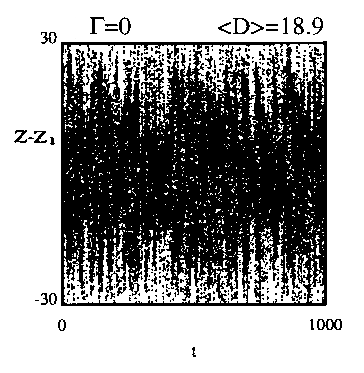}
 \end{minipage}
 \begin{minipage}{5.5in}
   \vspace{.3in}
   e) \includegraphics{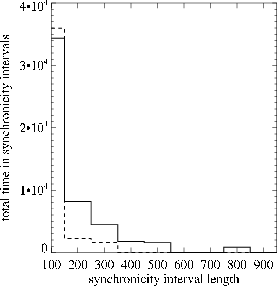}
 \end{minipage}
 \begin{minipage}{5.5in}
   \vspace{.2in}
 \end{minipage}
  \end{minipage}
   \caption{}
\end{figure}

\newpage
\begin{figure}
 \begin{minipage}{7in}
 \begin{minipage}{7in}
 \hspace{.3in} channel A \hspace{1.6in} channel B 
 \end{minipage}
 \begin{minipage}{7in}
 \hspace{0.2in} \raisebox{.5ex}{forcing} 
 \end{minipage}
 \begin{minipage}{7in}
 a) \resizebox{.3\textwidth}{!}{\includegraphics{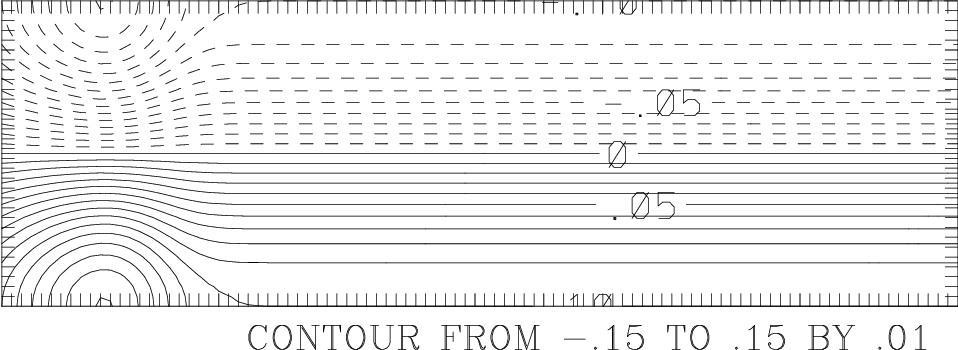}} \hspace{.2in}
   b) \resizebox{.3\textwidth}{!}{\includegraphics{kn15idframe2.eps}}
 \end{minipage}
 \begin{minipage}{7in}
  \bigskip \hspace{0.1in} \raisebox{.5ex}{n=0} 
 \end{minipage}
 \begin{minipage}{7in}
 c) \resizebox{.3\textwidth}{!}{\includegraphics{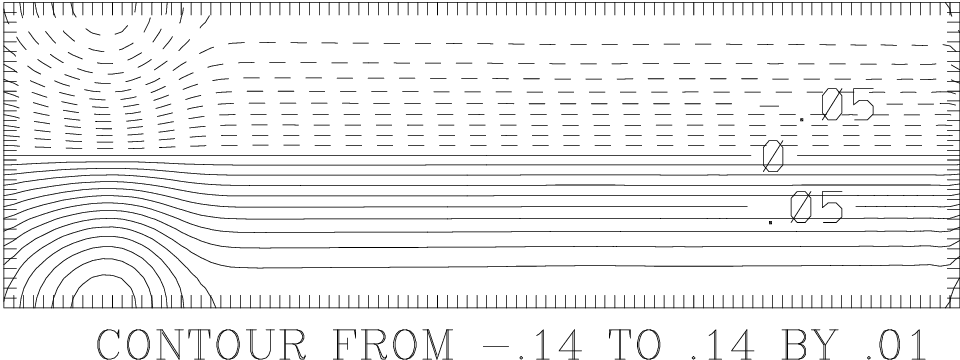}} \hspace{.2in}
   d) \resizebox{.3\textwidth}{!}{\includegraphics{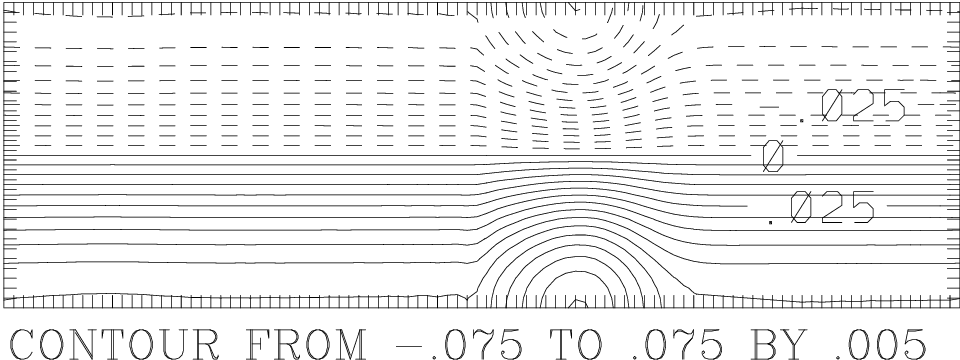}}
 \end{minipage}
 \begin{minipage}{7in}
  \bigskip \hspace{0.2in} \raisebox{.5ex}{n=2000} 
 \end{minipage}
 \begin{minipage}{7in}
 e) \resizebox{.3\textwidth}{!}{\includegraphics{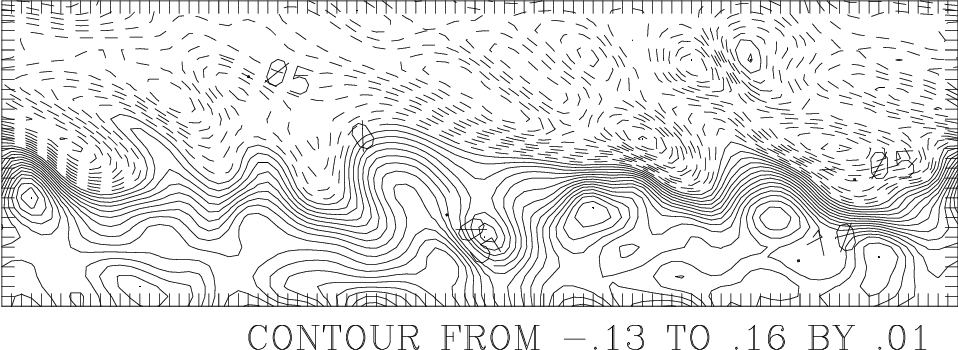}} \hspace{.2in}
   f) \resizebox{.3\textwidth}{!}{\includegraphics{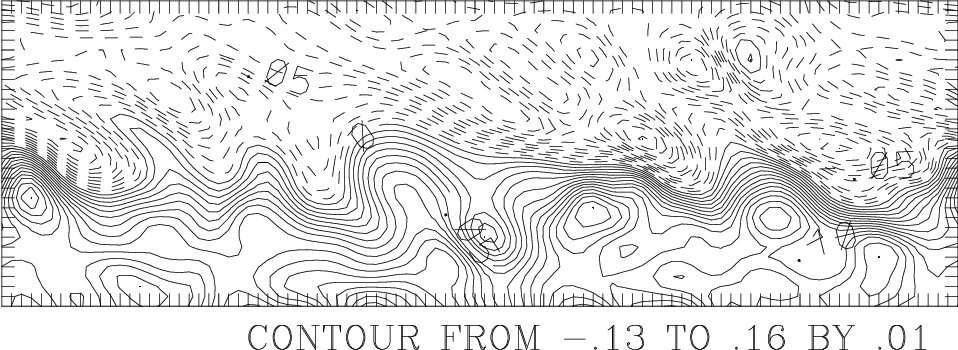}}
 \end{minipage}
  \begin{minipage}{7in}
     \vspace{.3in}
   \end{minipage}
  \end{minipage}
   \caption{}
\end{figure}

\newpage
\begin{figure}
\vspace{2in}
\begin{minipage}{7in}
\vspace{.3in}
a)\resizebox{.45\textwidth}{!}{\includegraphics{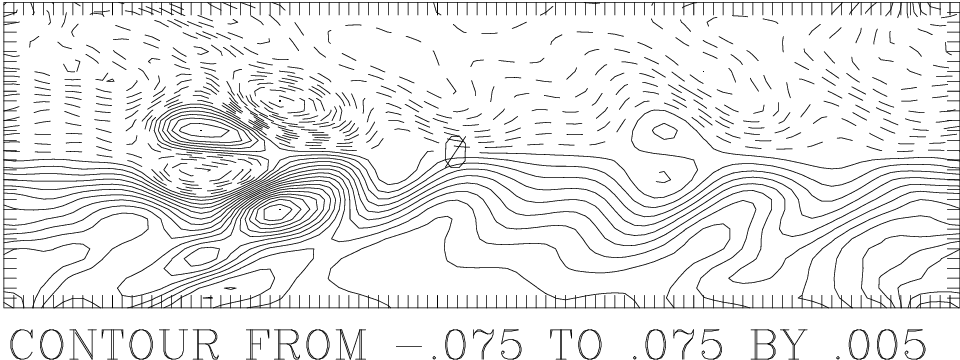}}
\end{minipage}
\begin{minipage}{7in}
\vspace{.3in}
b)\resizebox{.45\textwidth}{!}{\includegraphics{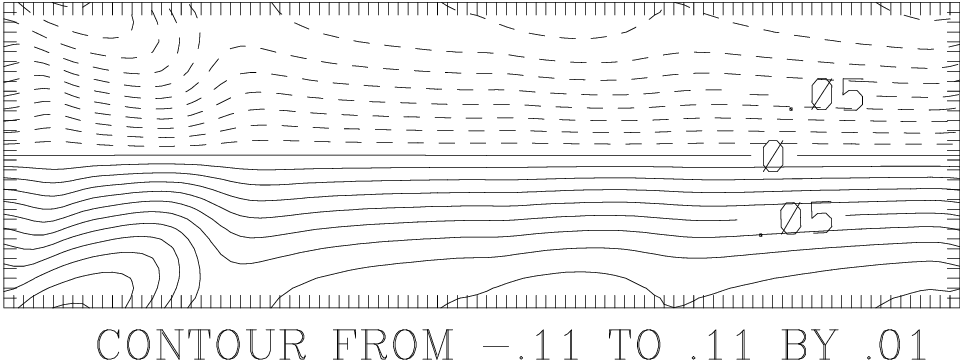}}
\end{minipage}
  \vspace{1in}
   \caption{}
\end{figure}

\newpage
\begin{figure}
\vspace{2in}
a)   \resizebox{.9\textwidth}{!}{\includegraphics{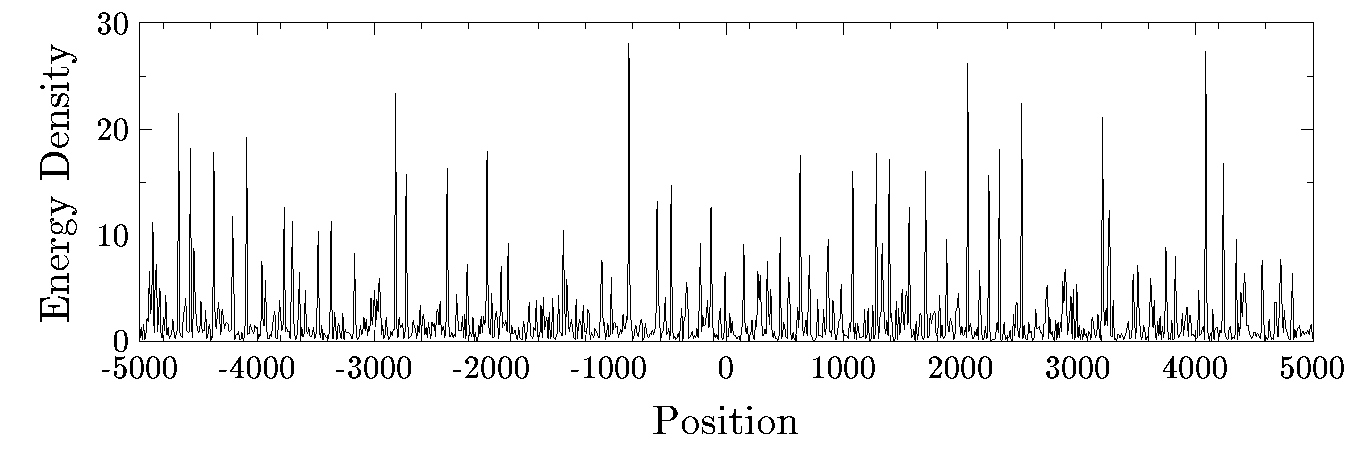}}\\ 
b)   \resizebox{.9\textwidth}{!}{\includegraphics{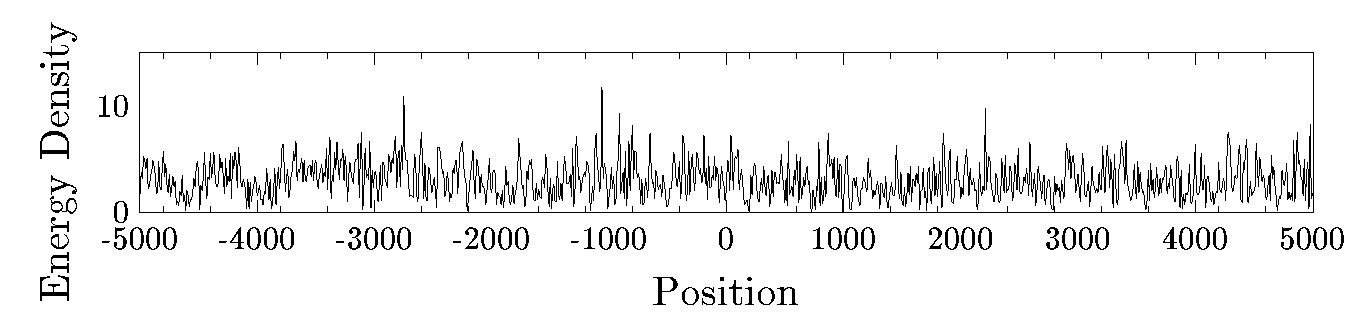}}
  \vspace{1in}
   \caption{}
\end{figure}

\newpage
\begin{figure}
\vspace{5in}
\resizebox{.4\textwidth}{!}{\includegraphics{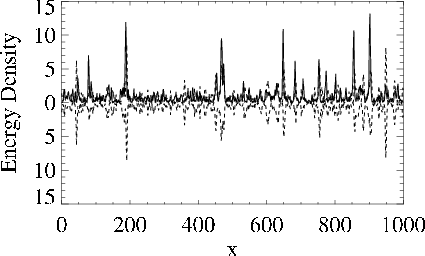}}
  \vspace{1in}
   \caption{}
\end{figure}

\newpage
\begin{figure}
\vspace{5in}
{{\resizebox{.4\textwidth}{!}{\includegraphics{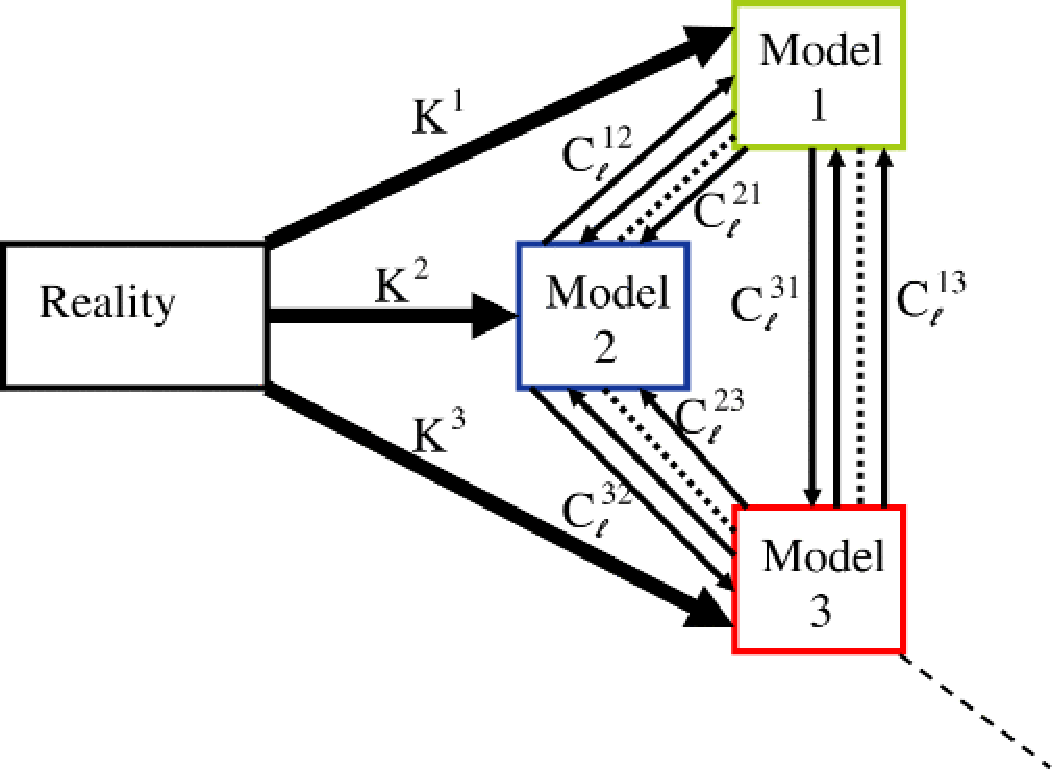}}}}, %
  \vspace{1in}
   \caption{}
\end{figure}

\newpage
\begin{figure}
\vspace{5in}
\resizebox{1.0\textwidth}{!}{\includegraphics{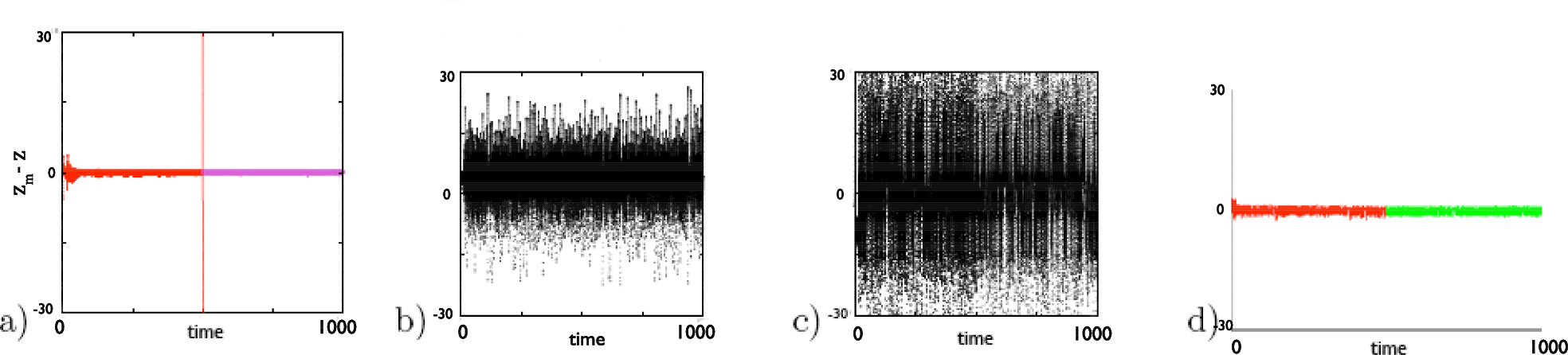}}
  \vspace{1in}
   \caption{}
\end{figure}

\end{document}